\documentclass[aps,prb,reprint,onecolumn,11pt,groupedaddress]{revtex4-2}
\usepackage[colorlinks=true,linkcolor=blue,citecolor=blue,urlcolor=blue]{hyperref}
\usepackage{amsmath}
\usepackage{amssymb} 
\usepackage[pdftex]{graphicx}
\usepackage{braket} 

\newcounter{firstbib}

\begin{document}


\title{Nuclear spin diffusion in the central spin system of a GaAs/AlGaAs quantum dot}


\author{Peter Millington-Hotze}
\affiliation{Department of Physics and Astronomy, University of
Sheffield, Sheffield S3 7RH, United Kingdom}
\author{Santanu Manna}
\affiliation{Institute of Semiconductor and Solid State Physics,
Johannes Kepler University Linz, Altenbergerstr. 69, 4040 Linz,
Austria}
\author{Saimon F. Covre da Silva}
\affiliation{Institute of Semiconductor and Solid State Physics,
Johannes Kepler University Linz, Altenbergerstr. 69, 4040 Linz,
Austria}
\author{Armando Rastelli}
\affiliation{Institute of Semiconductor and Solid State Physics,
Johannes Kepler University Linz, Altenbergerstr. 69, 4040 Linz,
Austria}
\author{Evgeny A. Chekhovich}
\email[]{e.chekhovich@sheffield.ac.uk} \affiliation{Department of
Physics and Astronomy, University of Sheffield, Sheffield S3 7RH,
United Kingdom}

\date{\today}

\begin{abstract}
The spin diffusion concept provides a classical description of a
purely quantum-mechanical evolution in inhomogeneously polarized
many-body systems such as nuclear spin lattices. The central spin
of a localized electron alters nuclear spin diffusion in a way
that is still poorly understood. In contrast to previous
predictions, we show experimentally that in GaAs/AlGaAs quantum
dots the electron spin accelerates nuclear spin diffusion, without
forming any Knight field gradient barrier. Such acceleration is
present even at high magnetic fields, which we explain as a result
of electron spin-flip fluctuations. Diffusion-limited nuclear spin
lifetimes range between 1 and 10~s, providing plenty of room for
recent proposals seeking to store and process quantum information
using quantum dot nuclear spins.
\end{abstract}


\maketitle

\section{Introduction}

\begin{figure}
\includegraphics[width=0.95\linewidth]{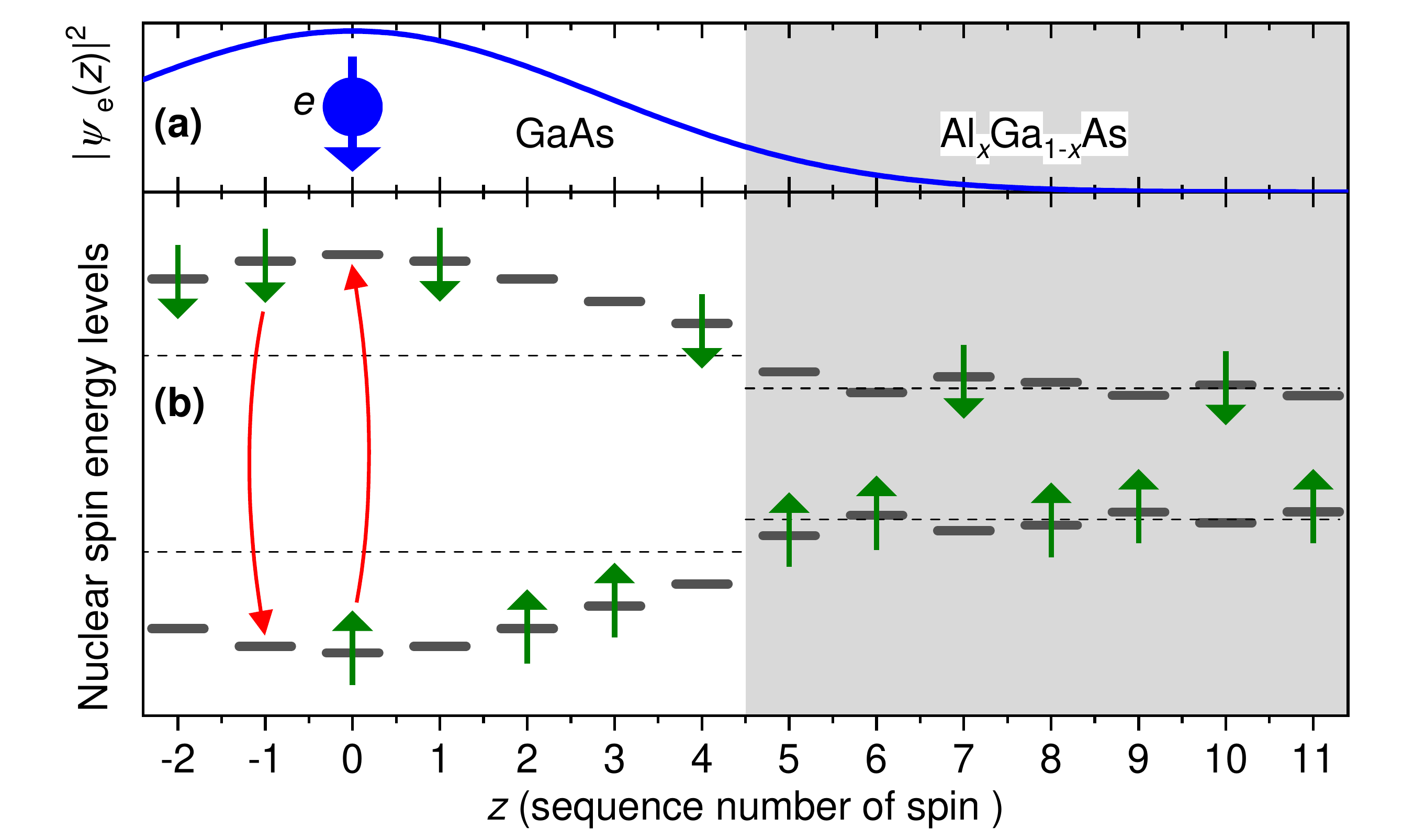}
\caption{\label{Fig:Intro} Schematic of a central spin model,
sketched for the one-dimensional case, along the growth axis $z$
of a GaAs/AlGaAs structure. (a) Wavefunction density
$|\psi_{\rm{e}}|^2$ of an electron ($e$, ball and arrow) localized
in GaAs. (b) Energy levels of the nuclei, that are depicted for
simplicity as spins 1/2, and can occupy states with $+1/2$ and
$-1/2$ spin projections (up and down arrows). Dashed lines show
the bulk nuclear spin energies dominated by the external magnetic
field $B_{\rm{z}}$. These bulk energies are generally different in
GaAs ($z\leq4$) and AlGaAs ($z\geq5$) due to the difference in
chemical shifts and homogeneous strain. The energies of the
individual nuclei are further shifted by the electron Knight field
(mainly in GaAs) and by the atomic-scale strain disorder in the
AlGaAs alloy. Magnetic dipole interaction between the nuclei can
result in spin exchange via a flip-flop process, sketched by the
curved arrows for nuclei at $z=-1$ and $z=0$ as an example. If
energy mismatch is larger than the nuclear spin level homogenous
broadening, for example for nuclei at $z=4$ and $z=5$, spin
exchange becomes prohibited, suppressing nuclear spin diffusion.}
\end{figure}

Interacting many-body spin ensembles exhibit a variety of
phenomena such as phase transitions \citep{Oja1997,Kessler2012}
spin waves \citep{deGennes1963,Shiomi2019} and emergent
thermodynamics \citep{Dorner2012,Bardarson2012}. Spin diffusion
\citep{Bloembergen1949,Blumberg1960} is one of the earliest
studied phenomena, where unitary quantum-mechanical evolution
results in an irreversible dissipation of a localized spin
polarization that is well described by the classical diffusion
model. Pure spin diffusion in homogeneous solids has been observed
in a few notable examples \citep{Zhang1998,Eberhardt2007}.
However, most systems of interest are inhomogeneous by nature. In
particular, magnetic (hyperfine) interaction with the central spin
of a localized electron [Fig.~\ref{Fig:Intro}(a)] causes shifts
(known as the Knight shifts\cite{Knight1949,Urbaszek2013}) in the
nuclear spin energy levels [Fig.~\ref{Fig:Intro}(b)]. The
resulting nuclear spin dynamics are complicated, as observed in a
wide range of solid state impurities
\citep{Leppelmeier1968,Wolfe1973,Paget1982,Lu2006,Hayashi2008,Wittmann2018,Ashok2018}
and semiconductor nanostructures
\citep{Tycko1995,Bayot1997,Hayashi2008,Nikolaenko2009,Reilly2010,Makhonin2010,Sallen2014}.
It is still an open question whether the inhomogeneous Knight
shifts accelerate \citep{Klauser2008,Reilly2010,Gong2011} or
suppress
\citep{Deng2005,Lai2006,Lu2006,Ramanathan2008,Gong2011,Sallen2014}
spin diffusion between the nuclei. Resolving this dilemma is both
of fundamental interest and practical importance for the recent
proposals to use nuclear spins as quantum memories and registers
\citep{Heshami2016,Denning2019,Chekhovich2020}, since spin
diffusion would set an ultimate limit to the longevity of any
useful quantum state.

Figure~\ref{Fig:Intro} sketches the central spin model where an
electron can be trapped in a GaAs layer surrounded by the AlGaAs
barriers, and for simplicity spin-1/2 particles are used to
describe the energy levels of the nuclei subject to the strong
external magnetic field $B_{\rm{z}}$. Any two nuclear spins $i$
and $j$ are coupled by the dipole-dipole interaction
$\propto2\hat{I}_{{\rm{z}},i}\hat{I}_{{\rm{z}},j}-(\hat{I}_{{\rm{x}},i}\hat{I}_{{\rm{x}},j}+\hat{I}_{{\rm{y}},i}\hat{I}_{{\rm{y}},j})$,
where $\hat{I}_{{\rm{x}},i}$, $\hat{I}_{{\rm{y}},i}$ and
$\hat{I}_{{\rm{z}},i}$ are the Cartesian components of the spin
operator ${\bf{I}}_i$ of the $i$-th nucleus. The last two terms of
the dipole-dipole interaction describe a flip-flop spin exchange
process (dashed arrows at $z=-1$ and $0$ in
Fig.~\ref{Fig:Intro}(b)), responsible for the transfer of spin
polarization in space, known as spin diffusion. The electric
quadrupolar moments of the spin-3/2 nuclei make them sensitive to
electric field gradients (EFGs), which can be induced by the
GaAs/AlGaAs interfaces roughness ($z=4.5$) or atomic-scale strains
arising from random positioning of the aluminium atoms
\citep{MALINOWSKI2001,Knijn2010} in AlGaAs ($z\ge5$). The
resulting energy splitting mismatch between the adjacent nuclei
can impede nuclear spin diffusion.

When an electron is added, its spin $\bf{s}$ couples to the
nuclear spin ensemble via hyperfine interaction:
\begin{align}
\mathcal{\hat{H}}_{\rm{hf}}=\sum_{j}{A_j(\hat{s}_{\rm{x}}\hat{I}_{{\rm{x}},j}+\hat{s}_{\rm{y}}\hat{I}_{{\rm{y}},j}+\hat{s}_{\rm{z}}\hat{I}_{{\rm{z}},j})},\label{Eq:Hhf}
\end{align}
where the summation goes over all nuclei $j$, and the coupling
constants $A_j$ are proportional to the electron density
$|\psi_{\rm{e}}({\bf{r}}_j)|^2$ at the nuclear sites ${\bf{r}}_j$.
On the one hand, through the term
$\hat{s}_{\rm{z}}\hat{I}_{{\rm{z}}}$, the electron spin can
produce a further diffusion barrier
\citep{Deng2005,Lai2006,Lu2006,Ramanathan2008,Gong2011,Sallen2014},
at the points of strong Knight shift gradient ($z=3$ in
Fig.~\ref{Fig:Intro}(a)). On the other hand, the electron spin can
mediate spin flip-flops between two distant nuclei with similar
energy splitting (e.g. $z=-2$ and $z=2$), potentially opening a
new channel for spin diffusion, especially at low magnetic fields
\citep{Klauser2008,Reilly2010,Gong2011}.

Here, we examine electron-controlled nuclear spin diffusion in
high quality epitaxial GaAs/AlGaAs quantum dots (QDs), which
emerged recently as an excellent platform for quantum light
emitters \citep{Liu2019,Zhai2020,Tomm2021} as well as spin qubits
\citep{Chekhovich2020,Zaporski2022} and quantum memories
\citep{Denning2019}. Crucially, we design experiments where
nuclear spin dynamics are examined either in absence or in
presence of the electron central spin, but under otherwise
identical initial nuclear spin state. In this way, we distinguish
with high accuracy the effects specific to the electron spin, and
demonstrate that no observable Knight field barrier is formed.
Instead, the nuclear-nuclear interactions mediated by the electron
spin accelerate nuclear spin diffusion up to unexpectedly high
magnetic fields -- we attribute this to the previously neglected
impact of the electron spin flips. Our results answer a
long-standing question in spin physics, and provide practical
guidelines for the design and optimization of quantum dot
electron-nuclear spin qubits and quantum memories.

\begin{figure*}
\includegraphics[width=0.95\linewidth]{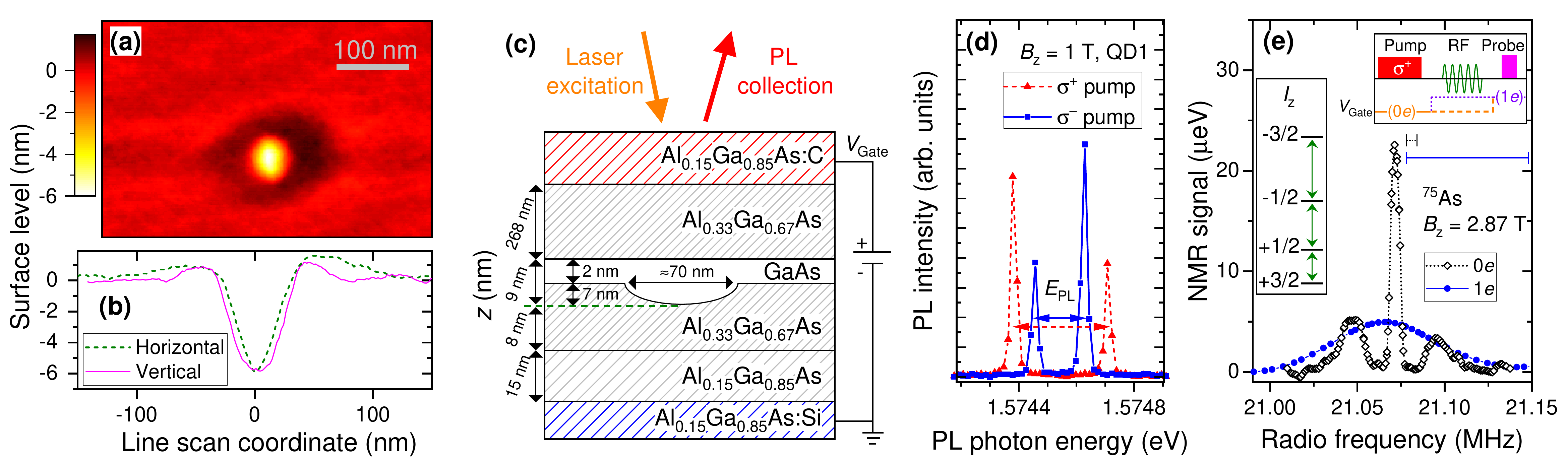}
\caption{\label{Fig:SampNMR} (a) Atomic force microscopy (AFM)
profile of the AlGaAs surface after nanohole etching. (b) Surface
level profiles taken along the horizontal and vertical lines
through the center of the nanohole in (a). (c) Schematic (not to
scale) of the sample structure. GaAs quantum dots (QDs) are formed
by infilling of the in-situ etched nanoholes in the bottom
Al$_{0.33}$Ga$_{0.67}$As barrier. The bottom (top)
Al$_{0.15}$Ga$_{0.85}$As layer is $n$ ($p$) type doped to form a
$p-i-n$ diode structure. External gate bias $V_{\rm{Gate}}$ is
applied for deterministic QD charging with electrons. (d)
Photoluminescence (PL) spectra of a negatively charged trion $X^-$
following $\sigma^+$ (triangles) and $\sigma^-$ optical pumping
which induces nuclear spin polarization that manifests in
hyperfine shifts of the Zeeman doublet spectral splitting $\Delta
E_{\rm{PL}}$. (e) Optically detected NMR of the $^{75}$As spin-3/2
nuclei measured in a single QD. Strain-induced quadrupolar shifts
of the nuclear spin-3/2 levels (left inset) give rise to an NMR
triplet with splitting $\nu_{\rm{Q}}\approx24$~kHz, observed in an
empty QD (0$e$, diamonds). Charging the QD with a single electron
(1$e$, circles) induces inhomogeneous Knight shifts observed as
NMR spectral broadening. The measurement is conducted using
``inverse NMR'' signal amplification technique
\citep{Chekhovich2012}, with spectral resolution shown by the
horizontal bars (smaller for 0$e$ and larger for 1$e$). The
measurement pump-probe cycle is shown in the top inset. The bias
$V_{\rm{Gate}}$ is tuned to 0$e$ charge state for optical pumping
of the nuclear spins and to 1$e$ state for their optical probing.
The radio frequency (RF) pulse is applied in the dark under either
0$e$ or 1$e$ bias.}
\end{figure*}

\section{Sample and experimental techniques}

The studied heterostructure is grown by in-situ etching of
nanoholes \cite{Heyn2009,Atkinson2012} in the AlGaAs surface
[Figs.~\ref{Fig:SampNMR}(a,b)], which are then infilled with GaAs
to form the quantum dots (QDs). The structure is processed into a
$p-i-n$ diode [Fig.~\ref{Fig:SampNMR}(c)] where an external bias
$V_{\rm{Gate}}$ is applied to charge QDs deterministically with
individual electrons (See details in Supplementary Section 1). In
this way, it is possible to study nuclear spin dynamics in an
empty (0$e$) or single-electron (1$e$) state. A static magnetic
field $B_{\rm{z}}$ is applied along the growth axis $z$ (Faraday
geometry) and the sample is kept at liquid helium temperature of
4.2~K. We use confocal microscopy configuration where QD
photoluminescence (PL) is excited and collected through an
aspheric lens with a focal distance of 1.45~mm and numerical
aperture of 0.58. The collected PL is dispersed in a two-stage
grating spectrometer, and recorded with a charge-coupled device
(CCD) camera.

The changes in the PL spectral splitting $\Delta E_{\rm{PL}}$ of a
negatively charged trion $X^-$ [see Fig.~\ref{Fig:SampNMR}(d)] is
the hyperfine shift $E_{\rm{hf}}$, which gives a measure of an
average nuclear spin polarization degree within the QD
\citep{Urbaszek2013}. The hyperfine shifts (also known as
Overhauser shifts) arise from the
$\hat{s}_{\rm{z}}\hat{I}_{{\rm{z}}}$ term of the hyperfine
interaction Hamiltonian (Eq.~\ref{Eq:Hhf}). Large nonequilibrium
nuclear spin polarization is generated on demand by exciting the
QD with a circularly polarized pump laser, which repeatedly
injects spin-polarized electrons into a QD, and causes nuclear
spin polarization build up via electron-nuclear spin flip-flops
described by the
$\hat{s}_{\rm{x}}\hat{I}_{{\rm{x}}}+\hat{s}_{\rm{y}}\hat{I}_{{\rm{y}}}$
part of Eq.~\ref{Eq:Hhf}. A small copper wire coil is placed near
the sample to produce radiofrequency (RF) oscillating magnetic
field perpendicular to the static magnetic field. Application of
the RF field allows for the energy spectrum of the nuclear spins
to be probed via nuclear magnetic resonance (NMR). Moreover, the
RF field can be used to depolarize the nuclear spins on-demand.
Further details can be found in the Supplementary Section 2,
including sample growth details, PL spectra, characterization of
QD charge state control, and additional results at an elevated
temperature of 15.2~K.

\section{Experimental results and discussion}

\subsection{Nuclear spin system of a GaAs quantum dot}

Figure~\ref{Fig:SampNMR}(e) shows NMR spectra of $^{75}$As in a
single GaAs QD, measured using ``inverse NMR'' technique with an
optical Pump-RF-Probe cycle shown in the top inset. For an empty
QD (open symbols), an NMR triplet is observed \citep{Ulhaq2016},
corresponding to the three magnetic-dipole transitions between the
four Zeeman-split states $I_{\rm{z}}=\{-3/2,-1/2,+1/2,+3/2\}$ of a
spin-3/2 nucleus (left inset). The central resolution-limited peak
originates from the $-1/2\leftrightarrow +1/2$ NMR transition that
is weakly affected by strain. The two satellite transition peaks
$\pm1/2\leftrightarrow \pm3/2$ are split from the central
transition peak by the strain-induced EFGs. The average splitting
$\nu_{\rm{Q}}\approx24$~kHz between the triplet components
corresponds to an average elastic strain of
$\approx2.6\times10^{-4}$ (Refs.
\citep{Chekhovich2018,Griffiths2019}). The satellite transitions
are inhomogeneously broadened, with non-zero NMR amplitudes
detected approximately in a range of $\nu_{\rm{Q}}\in[10,50]$~kHz,
indicating that elastic strain varies within the nanoscale volume
of the QD.

When a single electron occupies the QD, it induces inhomogeneous
Knight shifts that exceed the quadrupolar shifts, leading to a
broadened NMR peak [solid symbols in Fig.~\ref{Fig:SampNMR}(e)].
From the NMR peak width, the Knight shifts, characterizing the
typical coupling strength between the electron spin and an
individual nuclear spin, are estimated to be $A_j/h\approx50$~kHz,
where $h$ is the Planck's constant.

These NMR characterization results indicate a complex interplay of
dipolar, quadrupolar and hyperfine interactions governing the
nuclear spin dynamics, which we now investigate experimentally.

\subsection{Nuclear spin relaxation in a GaAs quantum dot}

\begin{figure*}
\includegraphics[width=0.95\linewidth]{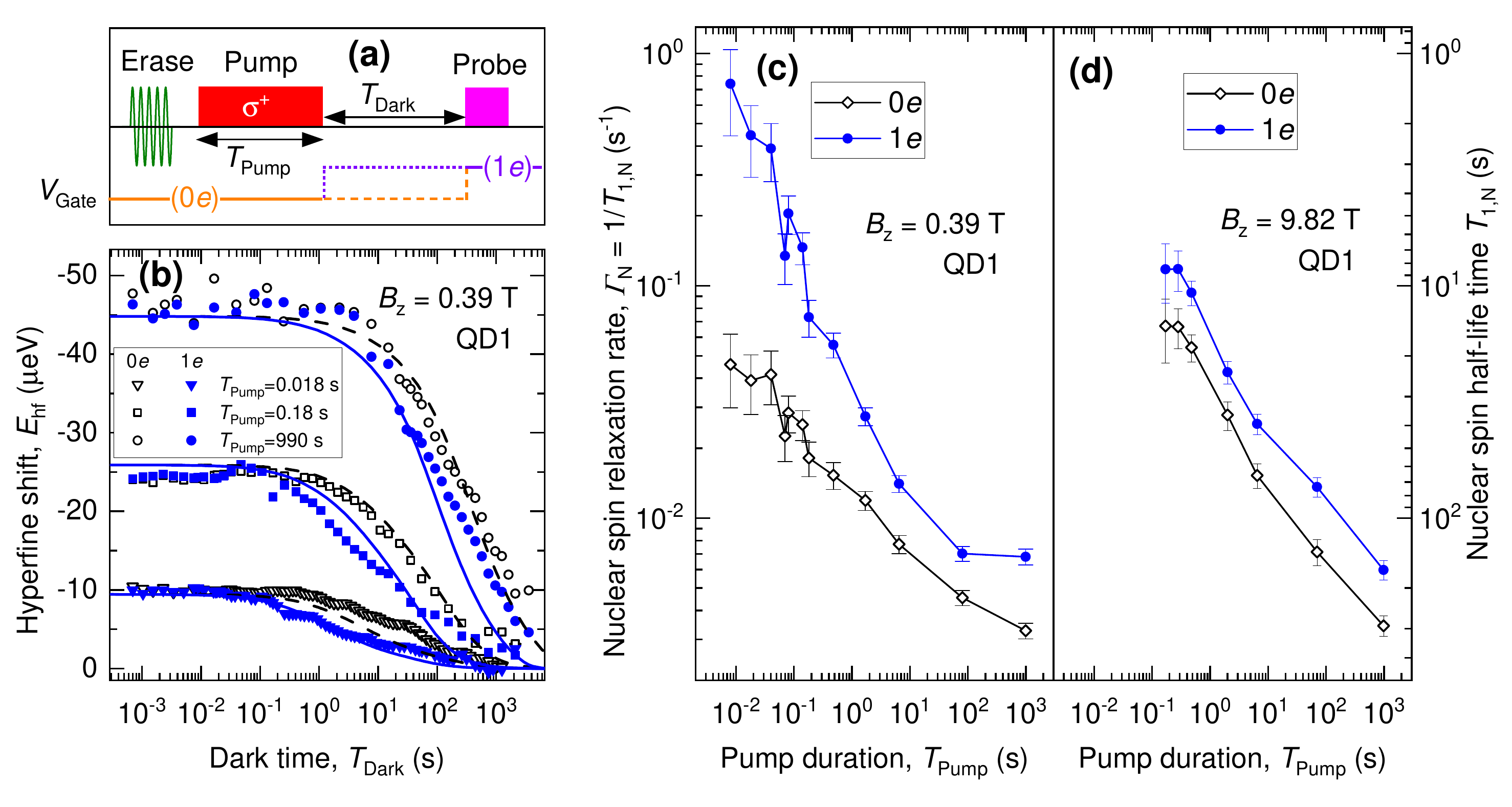}
\caption{\label{Fig:Diff} (a) NSR measurement cycle starting with
a radio frequency erase pulse, followed by circularly polarized
($\sigma^+$) optical pumping and then optical probing of the QD
nuclear spin polarization after dark evolution delay
$T_{\rm{Dark}}$ (see details in Supplementary Section 2). The
sample gate bias $V_{\rm{Gate}}$ is varied allowing to choose
between 0$e$ (dashed line) and 1$e$ (dotted line) QD charge state
during $T_{\rm{Dark}}$. (b) Dark time dependence of the hyperfine
shift $E_{\rm{hf}}$, which probes the average polarization of
$\approx10^5$ QD nuclear spins, weighted by the QD electron
density $|\psi_{\rm{e}}|^2$. The nuclear spin decay is measured
(symbols) at $B_{\rm{z}}=0.39$~T for different pumping times
$T_{\rm{Pump}}$ while keeping the QD empty (0$e$, open symbols) or
charged with one electron (1$e$, solid symbols) during the dark
time. Lines show numerical solution of the spin diffusion
Eq.~\ref{Eq:DiffEq}. (c) Fitted QD nuclear spin half-life times
$T_{\rm{1,N}}$ (right scale) and the corresponding NSR rates
$\varGamma_{\rm{N}}=1/T_{\rm{1,N}}$ (left scale) at magnetic field
$B_{\rm{z}}=0.39$~T. (d) same as (c) for $B_{\rm{z}}=9.82$~T. All
results are for the same individual dot QD1. Error bars are 95\%
confidence intervals.}
\end{figure*}

In the nuclear spin relaxation (NSR) experiment [see timing
diagram in Fig.~\ref{Fig:Diff}(a)] any remnant nuclear spin
polarization is first erased by saturating the $^{75}$As,
$^{69}$Ga and $^{71}$Ga NMR resonances in the entire
heterostructure \citep{Barrett1995}. This is followed by a
variable-duration ($T_{\rm{Pump}}$) optical pumping
\citep{Paget1982,Tycko1995,Hayashi2008,Nikolaenko2009} with photon
energies below the AlGaAs barrier bandgap, which prepares nuclear
spin polarization localized around the QD nanoscale volume. After
the pump laser is turned off, the gate bias $V_{\rm{Gate}}$ is set
to a desired level for a dark time $T_{\rm{Dark}}$ -- this way
evolution under 0$e$ or 1$e$ QD charge state is studied for
nominally identical initial nuclear spin polarizations. Finally,
the remaining polarization within the QD volume is probed through
an optically detected hyperfine shift $E_{\rm{hf}}$.

Figure~\ref{Fig:Diff}(b) shows the average QD nuclear spin
polarization as a function of the pump-probe delay $T_{\rm{Dark}}$
during which the sample is kept in the dark. The decay is
non-exponential, thus we characterize the NSR timescale
$T_{\rm{1,N}}$ by the half-life time over which the QD hyperfine
shift $E_{\rm{hf}}$ decays to 1/2 of its initial value. The NSR
rate is then defined as $\varGamma_{\rm{N}}=1/T_{\rm{1,N}}$. When
the pumping time $T_{\rm{Pump}}$ is increased, $T_{\rm{1,N}}$
notably increases, as can be seen in Figs.~\ref{Fig:Diff}(c,d).
Such dependence of $T_{\rm{1,N}}$ on $T_{\rm{Pump}}$ is observed
both in an empty (0$e$) and charged (1$e$) QD states, and in a
wide range of magnetic fields.

\subsection{Nuclear spin diffusion}

In order to explain the results of Fig.~\ref{Fig:Diff}, we note
that nuclear spin dipole-dipole interactions conserve the nuclear
spin polarization for any magnetic field exceeding the dipolar
local field, typically $\lesssim1$~mT. Therefore, at high magnetic
field the decay of nuclear spin polarization can proceed via two
routes: either via spin-conserving diffusion to the surrounding
nuclei, or spin transfer to external degrees of freedom, including
quadrupolar coupling to lattice vibrations
\citep{McNeil1976,Lu2006} or a hyperfine interaction with a charge
spin \citep{Lu2006,Latta2011,Vladimirova2017,Gillard2021} that is
in turn coupled to the lattice or other charges. Spin diffusion
can only take place if the spatial profile of the initial nuclear
spin polarization is inhomogeneous. By contrast, direct
spin-lattice and hyperfine interactions have no explicit
dependence on the spin polarization spatial profile. Optical
pumping that is short compared to spin diffusion timescales
creates nuclear spin polarization localized to the QD volume and
the resulting short $T_{\rm{1,N}}$ is therefore a clear indicator
of spin diffusion as the dominant NSR mechanism
\citep{Paget1982,Tycko1995,Hayashi2008,Nikolaenko2009}.
Conversely, long pumping provides enough time for nuclear
polarization to diffuse from the QD into the surrounding AlGaAs
barriers, suppressing any subsequent spin diffusion out of the QD
and increasing $T_{\rm{1,N}}$, as observed in
Figs.~\ref{Fig:Diff}(c,d).

In order to complement our experimental investigation we model the
spatiotemporal evolution of the nuclear spin polarization degree
$P _{\rm{N}}(t,z)$ by solving numerically the one dimensional spin
diffusion equation
\begin{align}
\frac{\partial P_{\rm{N}}(t,z)}{\partial
t}=D(t)\frac{{\partial}^{2} P _{\rm{N}}(t,z)}{\partial z^2} +\nonumber\\
+w(t)|\psi_{\rm{e}}(z)|^2 (P _{\rm{N},0}-P
_{\rm{N}}(t,z)),\label{Eq:DiffEq}
\end{align}
where the last term describes optical nuclear spin pumping with a
rate proportional to electron density $|\psi_{\rm{e}}(z)|^2$ and
the time-dependent factor $w(t)$ equal to 0 or $w_0$ when optical
pumping is off or on, respectively. Correspondingly, the spin
diffusion coefficient $D(t)$ takes two discrete values
$D_{\rm{Dark}}$ or $D_{\rm{Pump}}$ when optical pumping is off or
on, respectively. $P_{\rm{N},0}$ is a steady state nuclear spin
polarization degree that optical pumping would generate in the
absence of spin diffusion. Eq.~\ref{Eq:DiffEq} is solved
numerically and the parameters such as $D^{(ne)}_{\rm{Dark}}$,
$w_0(B_{\rm{z}})$, $D_{\rm{Pump}}(B_{\rm{z}})$ are varied to
achieve the best fit to the entire experimental datasets of
$E_{\rm{hf}}(T_{\rm{Pump}},T_{\rm{Dark}})$ measured at
$B_{\rm{z}}=0.39, 9.82$~T for empty ($n=0$) and charged ($n=1$) QD
states. The best-fit calculated dynamics are shown by the lines in
Fig.~\ref{Fig:Diff}(b) and capture well the main features of the
experimentally measured nuclear spin decay, confirming the
validity of the spin diffusion picture. The one-dimensional
character of diffusion, occurring predominantly along the sample
growth $z$ direction, is justified by the large ratio of the QD
diameter $\approx70$~nm to QD height $<9~$nm, and is further
verified by modeling two-dimensional spin diffusion (see
Supplementary Section 4).

\subsection{Effect of central spin on nuclear spin diffusion}

\begin{figure}
\includegraphics[width=0.95\linewidth]{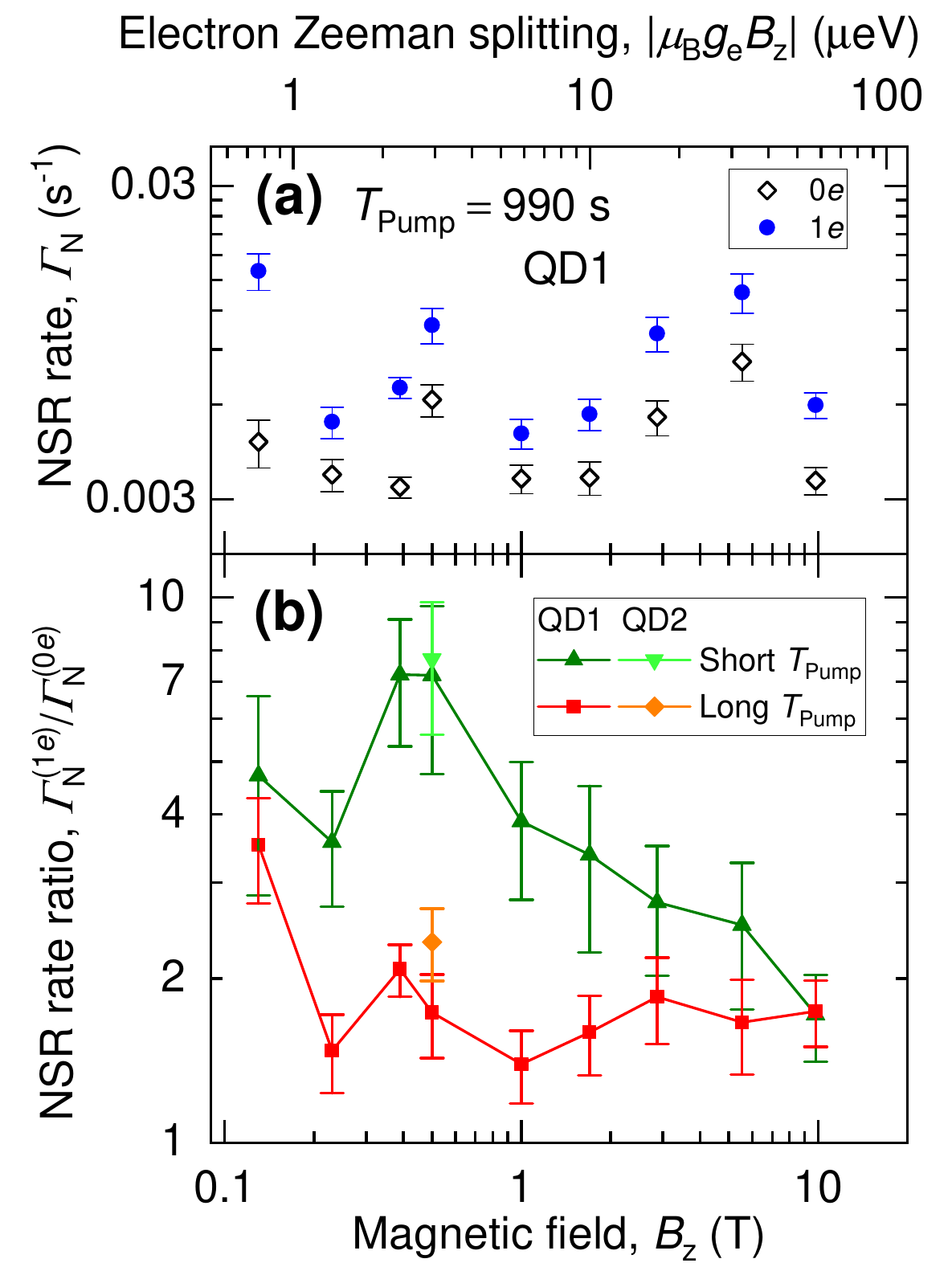}
\caption{\label{Fig:DiffBz} (a) Nuclear spin relaxation (NSR) rate
$\varGamma_{\rm{N}}$ as a function of $B_{\rm{z}}$ measured in
0$e$ (open symbols) and 1$e$ (solid symbols) states upon long
pumping $T_{\rm{Pump}}=990$~s. Top horizontal axis shows the
electron Zeeman splitting at zero nuclear spin polarization. (b)
Ratio $\varGamma_{\rm{N}}^{(1e)}/\varGamma_{\rm{N}}^{(0e)}$ of the
NSR rates in 1$e$ and 0$e$ charge states as a function of
$B_{\rm{z}}$ measured under long pumping $T_{\rm{Pump}}=990$~s
(squares) and short pumping $T_{\rm{Pump}}\in[0.08,0.48]$~s
(triangles). All results are for the same individual dot QD1,
except for the additional data from QD2 at $B_{\rm{z}}=0.5$~T in
(b). Error bars are 95\% confidence intervals.}
\end{figure}

Dividing the typical Knight shift of $\approx50$~kHz by half the
QD thickness (4.5~nm) we calculate the gradient and roughly
estimate the Knight shift difference of $\approx4.4$~kHz for the
two nearest-neighbor spins of the same isotope separated by
$a_0/\sqrt{2}$, where $a_0=0.565$~nm is the lattice constant. Such
difference significantly exceeds the energy that can be exchanged
with the dipole-dipole reservoir for a spin flip-flop to happen
\citep{Sallen2014} (the dipole-dipole energy is on the order of
$\approx h/T_{2,{\rm{N}}}$, where $T_{2,{\rm{N}}}\in[1,5]$~ms is
the nuclear spin-echo coherence time
\citep{Chekhovich2015,Chekhovich2020}). Therefore, one may naively
expect a Knight field gradient barrier to form and suppress spin
diffusion in an electron-charged QD (since the flip-flops would be
limited to the few nuclear spin pairs whose vector differences are
nearly orthogonal to the Knight field gradient). By contrast,
Figs.~\ref{Fig:Diff}(c,d) show that in experiment the NSR is
faster when the QD is occupied by a single electron (1$e$, solid
symbols) for all studied $T_{\rm{Pump}}$, demonstrating that no
significant Knight field barrier is formed. However, in order to
quantify the effect of the central spin on nuclear spin diffusion
we must distinguish it from other non-diffusion NSR mechanisms
introduced by the electron spin. To this end, we examine the
magnetic field dependence shown in Fig.~\ref{Fig:DiffBz}.

First, we examine a case where long optical pumping is used to
suppress spin diffusion and highlight the non-diffusion
mechanisms. Fig.~\ref{Fig:DiffBz}(a) shows the experimental
dependence $\varGamma_{\rm{N}}(B_{\rm{z}})$ for long
$T_{\rm{Pump}}=990$~s. In an empty QD (0$e$) spin diffusion is
still the dominant NSR mechanisms -- indeed, the observed rates
$\varGamma_{\rm{N}}^{(0e)}\in[3\times10^{-3},6\times10^{-3}]$~s$^{-1}$
are considerably higher than
$\varGamma_{\rm{N}}\in[6\times10^{-5},1\times10^{-3}]$~s$^{-1}$,
found in bulk crystals \citep{McNeil1976} such as semi-insulating
GaAs \citep{Lu2006}, where spin diffusion is negligible. The
electron-induced rates under long pumping
$\varGamma_{\rm{N}}^{(1e)}\in[4\times10^{-3},2\times10^{-2}]$~s$^{-1}$
are nearly independent of $B_{\rm{z}}$, and exceed the 0$e$ rates
by no more than a factor of
$\varGamma_{\rm{N}}^{(1e)}/\varGamma_{\rm{N}}^{(0e)}<4$ [squares
in Fig.~\ref{Fig:DiffBz}(b)]. Such small effect of the electron is
explained by the small strain of the GaAs/AlGaAs structures, which
reduces the efficiency of the non-diffusion NSR mechanisms related
to phonon and electron cotunneling. This is in stark contrast to
the large magnetic-field-induced variation
$\varGamma_{\rm{N}}^{(1e)}\in[5\times10^{-4},1\times10^{1}]$~s$^{-1}$
in Stranski-Krastanov self-assembled InGaAs QDs
\citep{Gillard2021}, where phonon and cotunneling mechanisms
dominate, both enabled by the noncollinear hyperfine interaction
\citep{Latta2011,Gillard2021}, arising in turn from the large
strain-induced nuclear quadrupolar shifts. Overall, the absolute
QD NSR rates shown in Fig.~\ref{Fig:DiffBz}(a) are nearly
constant, with some residual irregular dependence on magnetic
field which we ascribe to uncontrollable parameters, such as
charge state of the nearby impurities, or the initial spatial
profile of the nuclear spin polarization defined by the optical
nuclear spin pumping rate. By contrast, for any given $B_{\rm{z}}$
and $T_{\rm{Pump}}$ the ratio
$\varGamma_{\rm{N}}^{(1e)}/\varGamma_{\rm{N}}^{(0e)}$ shown in
Fig.~\ref{Fig:DiffBz}(b) gives a reliable measure, which we use to
examine the electron spin's effect on spin diffusion.

In order to discriminate the diffusion-related effect of the QD
electron spin, in addition to the long-pumping measurements
[squares in Fig.~\ref{Fig:DiffBz}(b)], we choose for each magnetic
field a short pumping time, typically
$T_{\rm{Pump}}\in[0.08,0.48]$~s, that yields initial QD nuclear
spin polarization at $\approx1/2$ of the steady-state long-pumping
polarization. The resulting short-pumping ratio
$\varGamma_{\rm{N}}^{(1e)}/\varGamma_{\rm{N}}^{(0e)}$ is shown by
the triangles in Fig.~\ref{Fig:DiffBz}(b) -- its excess over the
long-pumping ratio
$\varGamma_{\rm{N}}^{(1e)}/\varGamma_{\rm{N}}^{(0e)}$ is ascribed
to spin diffusion alone, discriminating it from any non-diffusion
mechanisms introduced by the electron spin. The
electron-spin-induced acceleration of the nuclear spin diffusion
is seen to be particularly pronounced at low magnetic fields
$B_{\rm{z}}\lesssim0.5$~T, consistent with the influence of the
electron-mediated nuclear-nuclear spin interaction
\citep{Klauser2008,Reilly2010,Gong2011}. Such pairwise indirect
interaction of nuclei $j$ and $k$ is derived from second order
perturbation expansion of Eq.~\ref{Eq:Hhf}:
\begin{align}
\mathcal{H}_{\rm{hf},j,k}^{\rm{ind}}\propto \frac{A_j A_k}{\Delta
E_{\rm{e}}}\hat{s}_{\rm{z}}\hat{I}^{(+)}_{j}\hat{I}^{(-)}_{k},\label{Eq:HhfInd}
\end{align}
where $\hat{I}_{j}^{(\pm)}=\hat{I}_{{\rm{x}},j}\pm
i\hat{I}_{{\rm{y}},j}$ and $\Delta
E_{\rm{e}}=\mu_{\rm{B}}g_{\rm{e}}B_{\rm{z}}+E_{\rm{hf}}$ is the
electron spin splitting due to both the Zeeman effect and the
nuclear-spin-induced hyperfine shift $E_{\rm{hf}}$. In our
experiments both contributions are negative, so that any nuclear
spin polarization increases $|\Delta E_{\rm{e}}|$. The rate of the
indirect nuclear-nuclear spin flip-flops scales as $\propto\Delta
E_{\rm{e}}^{-2}$. Consequently, the resulting acceleration of
nuclear spin diffusion in gate-defined GaAs QDs was previously
found to be limited to the low fields $B<0.02 - 0.75~$T (Refs.
\citep{Reilly2010,Gong2011,Malinowski2017}). By contrast,
Fig.~\ref{Fig:DiffBz}(b) shows that such acceleration persists at
magnetic fields well above $B_{\rm{z}}\gtrsim2$~T, with short- and
long-pumping $\varGamma_{\rm{N}}^{(1e)}/\varGamma_{\rm{N}}^{(0e)}$
ratios converging only at the maximum field $B_{\rm{z}}=9.82$~T.

One contributing factor is the smaller electron $g$-factor
$g_{\rm{e}}\approx-0.1$ (see Supplementary Section 2) and an order
of magnitude smaller number of nuclei in the studied epitaxial
QDs, which result in a smaller $|\Delta E_{\rm{e}}|$ and larger
$A_j$, respectively, when compared to the gate-defined QDs with
$g_{\rm{e}}\approx-0.4$. While these factors lead to a stronger
hyperfine-mediated couplings in the studied QDs, they do not
explain the magnetic field dependence: At high field
$B_{\rm{z}}=9.82$~T the electron spin Zeeman splitting is $|\Delta
E_{\rm{e}}|\approx58~\mu$eV, whereas at low field
$B_{\rm{z}}=0.39$~T we take into account both the Zeeman splitting
$\approx-2.3~\mu$eV and the time-averaged hyperfine shift
$E_{\rm{hf}}\approx-2.5~\mu$eV (half of the initial
$E_{\rm{hf}}\approx-5~\mu$eV under the shortest used
$T_{\rm{Pump}}\approx8~$ms) to estimate $|\Delta
E_{\rm{e}}|\approx5~\mu$eV. This would correspond to a factor of
$(58/5)^2\approx130$ reduction in the hyperfine mediated rates,
while the measured short-pumping-limit NSR rate reduces only by a
factor of $\approx6$ from
$\varGamma_{\rm{N}}^{(1e)}\approx0.74~{\rm{s}}^{-1}$ at low field
[Fig.~\ref{Fig:Diff}(c)] to
$\varGamma_{\rm{N}}^{(1e)}\approx0.12~{\rm{s}}^{-1}$ at high field
[Fig.~\ref{Fig:Diff}(d)]. Prompted by these observations, we point
out that Eq.~\ref{Eq:HhfInd} treats electron spin as isolated,
while in a real system the electron is coupled to external
environments such as phonons and other charges. A fluctuating
electron spin can accelerate nuclear spin diffusion, provided
there is a frequency component in the time-dependent Knight field
that equals the energy mismatch of a pair of nuclei
\citep{Khutsishvili1966,Horvitz1970} -- this contribution has been
considered for deep impurities \citep{Wolfe1973}, but has been
previously ignored in the context of III-V semiconductor
nanostructures. For similar GaAs/AlGaAs QDs \citep{Zhai2020} the
electron spin lifetime was reported to be 48~$\mu$s, while for
InGaAs/GaAs QDs with a similar tunnel coupling it was found to
vary with magnetic field between $\approx$50~$\mu$s and a few
milliseconds \citep{Gillard2021}. These lifetimes correspond to
fluctuation frequencies in a $[1,10]$~kHz range, indeed matching
the typical differences in the nuclear spin energies as revealed
by NMR spectra of Fig.~\ref{Fig:SampNMR}(e). Thus we speculate
that the intrinsic electron spin flips, governed e.g. by the
phonon relaxation and cotunneling coupling to the electron Fermi
reservoir of the $n$-doped layer
\citep{Kroutvar2004,Lu2010,Gillard2021}, contribute to
acceleration of nuclear spin diffusion in the studied GaAs QDs,
especially at high magnetic fields.

\subsection{Comparison with previous results on nuclear spin diffusion}

In order to understand what controls the rate of spin diffusion we
first make a comparison with Stranski-Krastanov InGaAs/GaAs and
InP/GaInP self-assembled QDs, where quadrupolar shifts are so
large (MHz range \citep{Bulutay2012,Chekhovich2012}) that all
nuclear spins are essentially isolated from each other,
eliminating spin diffusion and resulting in very long nuclear spin
lifetimes $T_{\rm{1,N}}^{(0e)}>10^4~$s in empty (0$e$) QDs
\citep{Lai2006,Greilich2007,Klauser2008,Chekhovich2010,Latta2011,Latta2011,Gillard2021}.
Even in presence of the electron spin (1$e$) the nuclear spin
diffusion takes place only inside the QD
\citep{Klauser2008,Latta2011}, without diffusion into the
surrounding material.

In the lattice-matched GaAs QDs the strain-induced effects are
smaller but not negligible, characterized by quadrupolar shifts
$\nu_{\rm{Q}}$ ranging approximately between 10 and 50~kHz within
the QD, as revealed by NMR spectra in Fig.~\ref{Fig:SampNMR}(e).
Nuclei in $I_{\rm{z}}=\pm1/2$ and $|I_{\rm{z}}|>1/2$ states must
be considered separately. The central transition between the
$I_{\rm{z}}=-1/2$ and $+1/2$ spin states is affected only by the
second order quadrupolar shifts, which scale as
$\propto\nu_{\rm{Q}}^2/\nu_{\rm{L}}$ and are within a few kHz for
the studied range of nuclear spin Larmor frequencies
$\nu_{\rm{L}}\in[1,130]$~MHz. These second order quadrupolar
shifts are comparable to the homogeneous nuclear spin linewidth
$\propto1/T_{2,{\rm{N}}}$, and therefore spin diffusion in
GaAs/AlGaAs QDs is expected to be unimpeded for the nuclei in the
$I_{\rm{z}}=\pm1/2$ states. By contrast, the $I_{\rm{z}}=\pm3/2$
spin states experience first order quadrupolar shifts
$\nu_{\rm{Q}}$, which are tens of kHz, significantly exceeding the
homogeneous NMR linewidths in the studied GaAs QDs. The resulting
dynamics of the $I_{\rm{z}}=\pm3/2$ nuclei is therefore sensitive
to nanoscale inhomogeneity of the strain-induced $\nu_{\rm{Q}}$.
From the NSR experiments [Fig.~\ref{Fig:Diff}(b)] we observe that
nuclear spin polarization relaxes to zero, even in an empty QD
(0$e$). This can only happen if spin diffusion is unimpeded not
only for the $I_{\rm{z}}=\pm1/2$ states, but also for the
$I_{\rm{z}}=\pm3/2$ states that are subject to the larger first
order quadrupolar shifts. Our interpretation is that strain in the
studied GaAs/AlGaAs QDs is a smooth function of spatial
coordinates: for nearly each QD nucleus it is possible to find
some neighboring nuclei with a strain variation small enough to
form a chain that conducts spin diffusion out of the GaAs QD into
the AlGaAs barriers.

Similarly fast NSR was observed previously in neutral QDs formed
by monolayer fluctuations in GaAs/AlGaAs quantum wells
\citep{Nikolaenko2009}. However, the opposite scenario was
realized in QDs with nanoholes etched in pure GaAs
\citep{Ulhaq2016} where nuclear spin polarization in an empty QD
(0$e$) was preserved for over $T_{1,{\rm{N}}}>5000$~s, suggesting
that some of the nuclei were frozen in the $I_{\rm{z}}=\pm3/2$
states, akin to quadrupolar blockade of spin diffusion in
self-assembled QDs. This contrast is rather remarkable since the
average strain, characterized by the average
$\nu_{\rm{Q}}\approx20 - 30$~kHz, is very similar for QDs grown in
nanoholes etched in AlGaAs (studied here) and in GaAs (Ref.
\citep{Ulhaq2016}). This comparison suggests that nuclear spin
dynamics are sensitive to QD morphology down to the atomic scale,
and could be affected by factors such as QD shape, as well as
GaAs/AlGaAs interface roughness and intermixing
\citep{Jusserand1992,SaherHelmy1997,Braun1997}. One possible
contributing factor is the QD growth temperature, which was
610$^\circ$~C in the structures used here, considerably higher
than 520$^\circ$~C in the structures studied previously
\citep{Atkinson2012,Ulhaq2016}. Further work would be required to
elucidate the role of all the underlying growth parameters.
Conversely, NSR can be a sensitive probe of the QD internal
structure.


We now quantify the spin diffusion process and compare our results
to the earlier studies in GaAs-based structures. The best fit of
the experimental NSR dynamics [lines in Fig.~\ref{Fig:Diff}(b)]
yields $D^{(0e)}_{\rm{Dark}}=2.2^{+0.7}_{-0.5}$~nm$^2$~s$^{-1}$
for the diffusion coefficient in an empty QD and in the absence of
optical excitation, in agreement with  $D = 1.0 \pm
0.15$~nm$^2$~s$^{-1}$ measured previously for spin diffusion
between two GaAs quantum wells across an Al$_{0.35}$Ga$_{0.65}$As
barrier \citep{MALINOWSKI2001}. This is approximately an order of
magnitude smaller than the first-principle estimate
\citep{Lowe1967,Redfield1969,Butkevich1988} of
$D^{(0e)}_{\rm{Dark}}\approx19$~nm$^2$~s$^{-1}$ for bulk GaAs (see
Supplementary Section 3) and the $D = 15.0 \pm 7$~nm$^2$~s$^{-1}$
value measured in pure AlAs \citep{Nguyen2014}. The reduced
diffusion in the AlGaAs alloy can be explained by the quadrupolar
disorder, arising from the random positioning of the aluminium
atoms \citep{MALINOWSKI2001}. Charging of the QD with a single
electron accelerates spin diffusion: we find
$D^{(1e)}_{\rm{Dark}}(9.82~{\rm{T}})=4.7^{+1.2}_{-1.0}$~nm$^2$~s$^{-1}$,
which increases to
$D^{(1e)}_{\rm{Dark}}(0.39~{\rm{T}})=7.7\pm1.9$~nm$^2$~s$^{-1}$ at
low magnetic fields where hyperfine-mediated nuclear-nuclear spin
exchange is enhanced in accordance with Eq.~\ref{Eq:HhfInd}. While
experimental data can be well described by the spin diffusion
Eq.~\ref{Eq:DiffEq}, it is worth noting the limited nature of the
model, which ignores the spatial variations of the nuclear-nuclear
couplings, the dependence of the electrons spin splitting $\Delta
E_{\rm{e}}$ on the instantaneous nuclear spin polarization, the
isotopic difference between $^{75}$As, $^{69}$Ga and $^{71}$Ga, as
well as neglecting any spin diffusion orthogonal to the sample
growth $z$ direction. As such, the diffusion coefficients $D$
should be treated as a coarse-grained description, aggregating the
numerous lattice-constant-scale parameters of the many-body spin
ensemble evolution.

\section{Discussion and Outlook}

The GaAs/AlGaAs QDs grown by nanohole infilling combine excellent
optical properties with low intrinsic strain, allowing for nuclear
spin qubit and quantum memory designs
\citep{Denning2019,Chekhovich2020,Zaporski2022}. The key
performance characteristic is the nuclear spin coherence time,
which can be extended up to $T_{2,{\rm{N}}}\approx10~$ms
(Ref.~\citep{Chekhovich2020}), but is ultimately limited by the
longitudinal relaxation time $T_{\rm{1,N}}$. Moreover, it is the
state longevity of the nuclei interfaced with the QD electron spin
that is relevant. Thus, one should consider the NSR time in the
regime of short pumping, found here to range from
$T_{\rm{1,N}}^{(1e)}\approx1~$s at low magnetic fields to
$T_{\rm{1,N}}^{(1e)}\approx10~$s at high fields. For nuclear spin
quantum computing with the typical 10~$\mu$s coherent control
gates \citep{Chekhovich2020}, a large number of operations
$\gtrsim 10^5$ would be possible without the disruptive effect of
spin diffusion.

In conclusion, we have addressed the long-standing dilemma of
whether the central spin of an electron accelerates or suppresses
diffusion in a nuclear spin lattice. We have used
variable-duration optical pumping
\citep{Paget1982,Tycko1995,Hayashi2008,Nikolaenko2009} to identify
nuclear spin diffusion as the dominant NSR mechanism. In contrast
to previous studies of nuclear spin diffusion
\cite{Paget1982,Lu2006,Tycko1995,Bayot1997,Makhonin2010,Sallen2014},
we use a charge tunable structure and probe nuclear spin dynamics
with and without the electron under otherwise identical conditions
-- importantly, our QD charge control is achieved without
reverting to optical pumping \citep{Makhonin2010,Sallen2014}, thus
eliminating the unwanted charge fluctuations. Combining these two
aspects, we conclude that in a technologically important class of
lattice matched GaAs/AlGaAs nanostructures the electron spin
accelerates the nuclear spin diffusion, with no signature of a
Knight-field-gradient barrier. We expect these findings to be
relevant for a range of lattice-matched QDs
\citep{Nikolaenko2009,Reilly2010,Gong2011,Malinowski2017} and
shallow impurities \citep{Lu2006}, whereas an efficient spin
diffusion barrier can arise from an electron with sub-nanometer
localization \citep{Wolfe1973}. Future work can examine reduction
of spin diffusion in low-strain nanostructures. The proximity of
the $n$-doped layer, acting as a sink for nuclear polarization, as
well as QD morphology can be optimized. Alternatively, pure AlAs
barriers can be used to grow GaAs QDs with well isolated Ga
nuclei, potentially offering long-lived spin memories and qubits.

\begin{acknowledgments}
Acknowledgements: P.M-H. and E.A.C. were supported by EPSRC
through a doctoral training grant and EP/V048333/1, respectively.
E.A.C. was supported by a Royal Society University Research
Fellowship. A.R. acknowledges support of the Austrian Science Fund
(FWF) via the Research Group FG5, I 4320, I 4380, I 3762, the
European Union's Horizon 2020 research and innovation program
under Grant Agreements No. 899814 (Qurope) and No. 871130
(Ascent+), the Linz Institute of Technology (LIT), and the LIT
Secure and Correct Systems Lab, supported by the State of Upper
Austria. E.A.C is grateful to Ren\'{e} Dost for advice on sample
processing. Author contributions: S.M., S.F.C.S and A.R.
developed, grew and processed the quantum dot samples. P.M-H. and
E.A.C. conducted the experiments. E.A.C. drafted the manuscript
with input from all authors. E.A.C. coordinated the project.
\end{acknowledgments}


\newcommand{\RedText}[1]{{#1}}
\renewcommand{\thesection}{Supplementary Section \arabic{section}}
\setcounter{section}{0}
\renewcommand{\thefigure}{\arabic{figure}}
\renewcommand{\figurename}{Supplementary Figure}
\renewcommand{\theequation}{S\arabic{equation}}
\renewcommand{\thetable}{\arabic{table}}
\renewcommand{\tablename}{Supplementary Table}

\makeatletter
\def\l@subsection#1#2{}
\def\l@subsubsection#1#2{}
\makeatother

\pagebreak \pagenumbering{arabic}
\newpage


\section*{Supplementary Material}

\section{Sample structure}
\label{sec:Sample}

The sample is grown using molecular beam epitaxy (MBE) on a
semi-insulating GaAs (001) substrate. The growth starts with a
layer of Al$_{0.95}$Ga$_{0.05}$As followed by a single pair of
Al$_{0.2}$Ga$_{0.8}$As and Al$_{0.95}$Ga$_{0.05}$As layers acting
as a Bragg reflector in optical experiments. Then, a 95~nm thick
layer of Al$_{0.15}$Ga$_{0.85}$As is grown. The rest of the
structure follows the schematic shown in Fig.~2(c) of the main
text beginning with a 95~nm thick layer of
Al$_{0.15}$Ga$_{0.85}$As doped with Si at a volume concentration
of $1.0\times10^{18}$~cm$^{-3}$. The low Al concentration of
$0.15$ in the Si doped layer mitigates the issues caused by the
deep DX centers \citep{SupOshiyama1986,SupMooney1990,SupZhai2020}.
Under optical excitation this Al$_{0.15}$Ga$_{0.85}$As:Si gives
rise to broad photoluminescence at around 730~nm as observed in
Supplementary Fig.~\ref{Fig:SPL}(a). The $n$-type doped layer is
followed by the electron tunnel barrier layers: first a 15~nm
thick Al$_{0.15}$Ga$_{0.85}$As layer and then a 15~nm thick
Al$_{0.33}$Ga$_{0.67}$As layer. Aluminium droplets are grown on
the surface of the Al$_{0.33}$Ga$_{0.67}$As layer and are used to
etch the nanoholes \citep{SupHeyn2009,SupAtkinson2012}. An atomic
force microscopy (AFM) image of a similar sample in Fig.~2(a) of
the main text shows a typical nanohole with a depth of
$\approx6.5$~nm and $\approx70$~nm in diameter. Next, a 2.1~nm
thick layer of GaAs is grown to form QDs by infilling the
nanoholes as well as to form the quantum well (QW) layer. Thus,
the maximum height of the QDs in the growth $z$ direction is
$\approx9$~nm. Low temperature PL of QDs and QW is observed
[Supplementary Fig.~\ref{Fig:SPL}(a)] at 785~nm and 690~nm,
respectively. The GaAs layer is followed by a 268~nm thick
Al$_{0.33}$Ga$_{0.67}$As barrier layer. Finally, the $p$-type
contact layers doped with C are grown: a 65~nm thick layer of
Al$_{0.15}$Ga$_{0.85}$As with a $5\times10^{18}$~cm$^{-3}$ doping
concentration, followed by a 5~nm thick layer of
Al$_{0.15}$Ga$_{0.85}$As with a $9\times10^{18}$~cm$^{-3}$
concentration, and a 10~nm thick layer of GaAs with a
$9\times10^{18}$~cm$^{-3}$ concentration.

The sample is processed into a $p-i-n$ diode structure. Mesa
structures with a height of 250~nm are formed by etching away the
$p$-doped layers and depositing Ni(10 nm)/AuGe(150 nm)/Ni(40
nm)/Au(100 nm) on the etched areas. The sample is then annealed to
enable diffusion down to the $n$-doped layer to form the ohmic
back contact. The top gate contact is formed by depositing Ti(15
nm)/Au(100 nm) on to the $p$-type surface of the mesa areas. The
sample gate bias $V_{\rm{Gate}}$ is the bias of the $p$-type top
contact with respect to the grounded $n$-type back contact. By
changing $V_{\rm{Gate}}$ the equilibrium charge state of the
quantum dot is tuned using the Coulomb blockade effect. Due to the
large thickness of the top Al$_{0.33}$Ga$_{0.67}$As layer, the
tunneling of the holes is effectively blocked, whereas tunnel
coupling to the $n$-type layer enables deterministic charging of
the quantum dots with electrons.

\begin{figure}
\includegraphics[width=0.98\linewidth]{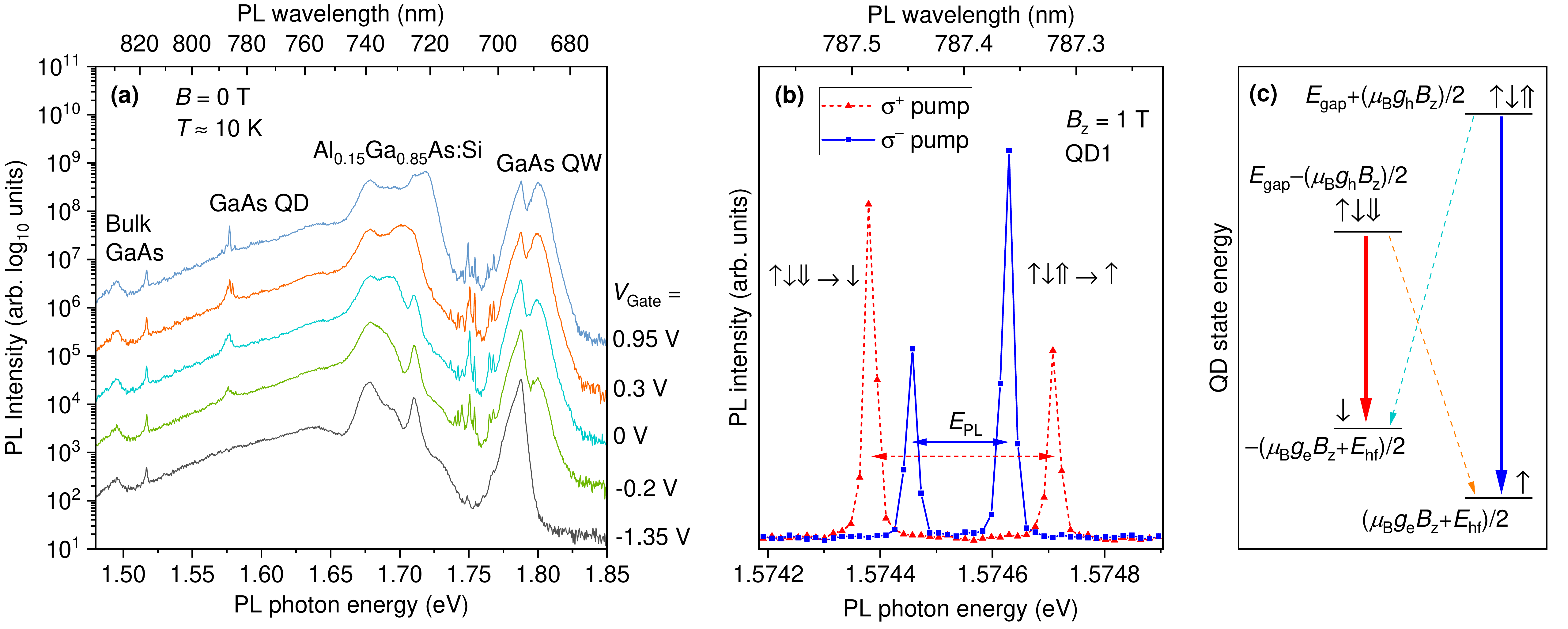}
\caption{\label{Fig:SPL} (a) Broad range photoluminescence (PL)
spectra measured under 532~nm laser excitation at different gate
biases $V_{\rm{Gate}}$. Spectra are offset in a vertical direction
(log scale) by a factor of 10 for clarity. Spectral features
arising from the different parts of the sample are labeled
accordingly. (b) High resolution PL spectra of a negatively
charged $X^-$ trion following $\sigma^+$ (triangles) and
$\sigma^-$ (squares) circularly polarized optical pumping, which
creates $s_{\rm{z}}=-1/2$ ($\downarrow$) and $s_{\rm{z}}=+1/2$
($\uparrow$) spin polarized electrons, respectively. The electrons
transfer their spin to the nuclei via magnetic (hyperfine)
interaction, resulting in a build up of negative or positive net
nuclear spin polarization, respectively. Through the same
hyperfine interaction, the average nuclear spin polarization
shifts the $s_{\rm{z}}=-1/2$ and $s_{\rm{z}}=+1/2$ electron spin
energy levels in the opposite directions. These (Overhauser)
shifts $E_{\rm{hf}}$ lead to the observed change in the spectral
splitting of the trion PL, where the two components of the doublet
correspond to an electron-hole recombination in presence of
another electron in a $s_{\rm{z}}=-1/2$ or $s_{\rm{z}}=+1/2$
state. (c) \RedText{Energy level diagram. The electron ground
state is split by the Zeeman energy
$\mu_{\rm{B}}g_{\rm{e}}B_{\rm{z}}$ and the hyperfine shift
$E_{\rm{hf}}$. The $X^-$ trion energy includes the QD bandgap
energy $E_{\rm{gap}}$ and the Zeeman splitting of the unpaired
hole with a positive ($\Uparrow$) or negative ($\Downarrow$)
momentum projection. The valence band hole hyperfine effect can be
neglected due to its smaller magnitude
\citep{SupChekhovich2013NatPhys}. The electron and hole
$g$-factors are $g_{\rm{e}}$ and $g_{\rm{h}}$, respectively, with
$|g_{\rm{h}}|\gg|g_{\rm{e}}|$ in the studied QDs. Solid arrows
depict the two optically allowed transitions responsible for the
spectral doublet in (b). The dashed lines show the two forbidden
``diagonal'' transitions.}}
\end{figure}

\section{Experimental details and additional results}
\label{sec:ExpTechn}

The sample is placed in a liquid helium bath cryostat. A
superconducting coil is used to apply magnetic field up to
$B_{\rm{z}}=10$~T. The field is parallel to the sample growth
direction and the optical axis $z$ (Faraday geometry). We use
confocal microscopy configuration. An aspheric lens with a focal
distance of 1.45~mm and NA=0.58 is used as an objective for
optical excitation of the QD and for photoluminescence (PL)
collection. \RedText{The excitation laser is focused into a spot
with a diameter of $\approx1~\mu$m.} The collected PL is dispersed
in a two-stage grating spectrometer, each stage with a 0.85~m
focal length, and recorded with a charge-coupled device (CCD)
camera. The changes in the spectral splitting of a negatively
charged trion $X^-$, derived from the PL spectra, are used to
measure the hyperfine shifts $E_{\rm{hf}}$ proportional to the
nuclear spin polarization degree.

Investigation of spin diffusion relies on the ability to prepare a
reproducible spatial distribution of the nuclear spin
polarization. This is achieved with a radiofrequency (RF) erase
pulse (Fig.~3(a) of the main text) which effectively resets the
nuclear spin polarization to zero in the entire sample by
saturating the nuclear magnetic resonance of the As and Ga
isotopes. The required oscillating magnetic field $B_{\rm{x}}\perp
z$ is produced by a coil placed at a distance of $\approx0.5$~mm
from the QD sample. The coil is made of 10 turns of a 0.1~mm
diameter enameled copper wire wound on a $\approx0.4$~mm diameter
spool in 5 layers, with 2 turns in each layer. The coil is driven
by a class-A RF amplifier (rated up to 20~W) which is fed by the
output of an arbitrary waveform generator. The spectrum of the RF
excitation consists of three bands, each 340~kHz wide and centered
on the NMR frequency of the corresponding As or Ga isotope. Each
band is generated as a frequency comb
\citep{SupChekhovich2013NatPhys} with a mode spacing of 120~Hz,
much smaller than the homogeneous NMR linewidth. The RF power
density in the comb is chosen to be low enough and the RF pulse
duration long enough (ranging between 0.1 and 10~s depending on
magnetic field) to achieve noncoherent exponential depolarization
of the nuclear spin ensemble.

Optical pumping of the QD nuclear spin polarization is achieved
using the emission of a 690~nm circularly polarized diode lasers,
which is resonant with the GaAs QW states, as seen in
Supplementary Fig.~\ref{Fig:SPL}(a). It is therefore possible that
optical pumping results in dynamical nuclear polarization not only
in the QDs but also in the adjacent parts of the QW. On the other
hand, the pump laser photon energy is well below the bandgap of
the AlGaAs barriers. For that reason we assume that dynamic
nuclear polarization in AlGaAs is induced only through spin
diffusion from the GaAs layer of the QW and QDs. During the pump
the sample gate is set to a large reverse bias, typically
$V_{\rm{Gate}}=-2$~V. The pump power is $\approx300~\mu$W, which
is two orders of magnitude higher than the ground-state PL
saturation power. The resulting hyperfine shifts do not exceed
$|E_{\rm{hf}}|<50~\mu$eV, corresponding to initial nuclear spin
polarization degree within $|P_{\rm{N},0}|\lesssim0.4$. While
polarization as high as $P_{\rm{N},0}\approx0.8$ is possible
\citep{SupChekhovich2017}, we deliberately use lower values to
ensure linear regime of spin diffusion, free from
hyperpolarization regime corrections \citep{SupWang2021}. For
optical probing of the nuclear spin polarization we use a diode
laser emitting at 640~nm. Sample forward bias, typically $+0.5$~V,
and the probe power are chosen to maximize (saturate) PL intensity
of the ground state $X^-$ trion. The duration of the probe pulse,
typically 20~ms, is selected to ensure minimal (few percent)
depolarization of the nuclear spins during probing. Supplementary
Fig.~\ref{Fig:SPL}(b) shows $X^-$ PL probe spectra measured at
$B_{\rm{z}}=1$~T following optical pumping with $\sigma^+$
(triangles) and $\sigma^-$ (squares) circular polarization.
\RedText{The difference in spectral splitting of the $X^-$ trion
doublet reveals the hyperfine shifts $E_{\rm{hf}}$ [see energy
level diagram in Supplementary Fig.~\ref{Fig:SPL}(c)]. These
shifts are used to monitor the average QD nuclear spin
polarization in NSR experiments such as shown in Fig.~3(b) of the
main text.}

\RedText{The energy splitting of the two electron spin states
$\Delta E_{\rm{e}}$ is a sum of the Zeeman splitting
$\mu_{\rm{B}}g_{\rm{e}}B_{\rm{z}}$ (where $\mu_{\rm{B}}$ is the
Bohr magneton) and the hyperfine splitting $E_{\rm{hf}}$, arising
from the nuclear spin polarization. We quantify the $g$-factor
$g_{\rm{e}}$ of a resident electron using photoluminescence
spectroscopy of a negatively charged trion. In Faraday geometry,
two out of four optical transitions are forbidden, so that only
the difference $g_{\rm{h}}-g_{\rm{e}}$ of the heavy hole and
electron $g$-factors can be accessed. In order to derive the
individual $g$-factors, we measure photoluminescence in oblique
field configuration, where the sample growth axis is tilted by
$\theta\approx12^\circ$ away from the static magnetic field. In
this configuration the ``diagonal'' transitions, shown by the
dashed lines in Supplementary Fig.~\ref{Fig:SPL}(c), become weakly
allowed. Owing to the nearly vanishing electron $g$-factor in this
type of GaAs/AlGaAs QDs \citep{SupUlhaq2016}, all four $X^-$
transitions can be resolved in our setup only in high magnetic
field $B=10$~T and in presence of the optically induced hyperfine
shifts. The experiment is conducted using an optical pump-probe
method. The probe PL spectra are shown in Supplementary
Fig.~\ref{Fig:Sge} as a function of the half-wave plate angle. The
angle is varied to control the degree of circular polarization of
the pump laser and the resulting hyperfine shift $E_{\rm{hf}}$.
The two weak transitions (labeled by the arrows) become visible
when the splitting of the two bright transitions is maximized by
the hyperfine shift. We further measure the spectral splitting of
the two bright transitions after RF depolarization of the nuclei,
which results in $E_{\rm{hf}}\approx0$. It is then possible to
perform linear fit of the PL energies of all four $X^-$
transitions and derive the $g$-factors. For QD1 studied in the
main text and examined in Supplementary Fig.~\ref{Fig:Sge} we find
the 95\% confidence estimate $g_{\rm{e}}\approx-0.101\pm0.007$ for
the $g$-factor of a single resident electron. From the hole spin
splitting of $X^-$ at $B=10$~T we estimate the hole $g$-factor in
presence of two electrons as $g_{\rm{h}}\approx+1.68$. This value
should be treated as a rough estimate because of the significant
nonlinearity in hole Zeeman splitting for this type of QDs
\citep{SupUlhaq2016}. We also measure the $g$-factors in a neutral
exciton $X^0$, using PL of the dark states: we find
$g_{\rm{e}}\approx-0.090\pm0.035$ for the electron in presence of
one hole. It is notable that the electron $g$-factor is nearly
unaffected by the extra hole \citep{SupHuber2019}. Using PL
spectroscopy of the $X^-$ trion state, we have measured
$g$-factors in two more QDs from the same sample to find
$g_{\rm{e}}\approx-0.077\pm0.018$ and
$g_{\rm{e}}\approx-0.107\pm0.002$ for a resident electron in QD2
and QD3, respectively. From the $X^0$ PL of QD2 we find
$g_{\rm{e}}\approx-0.12\pm0.01$ for an electron in presence of a
hole, whereas no dark excitons could be observed in QD3. The
$g$-factors found here are in good agreement with the previous
studies on the samples where QDs were grown in nanoholes etched in
pure GaAs \citep{SupUlhaq2016}.}

\begin{figure}
\includegraphics[width=0.7\linewidth]{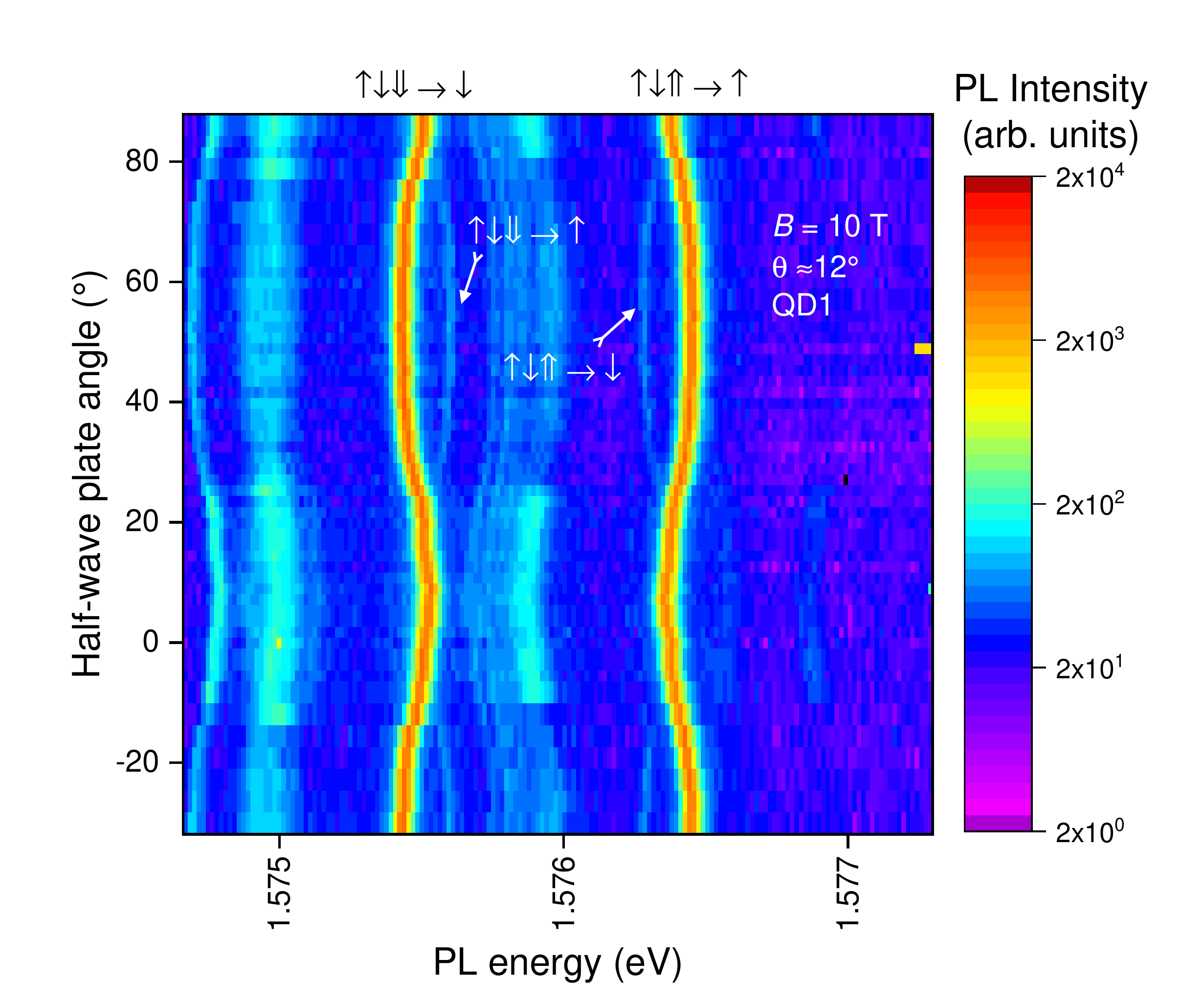}
\caption{\label{Fig:Sge} \RedText{Photoluminescence spectra of a
negatively charged trion $X^-$ measured in oblique magnetic field
$B=10$~T tilted by $\theta\approx12^\circ$ from the Faraday
geometry. The measurement uses a pump-probe protocol, where the
angle of a half-wave plate on the pump laser is varied, while the
probe laser is used to detect the resulting changes in PL
spectrum. The two bright lines correspond to the two allowed
transitions. When the circularly polarized pump generates a
sufficiently large hyperfine shift, the two weakly allowed trion
transitions, labeled by the arrows, become resolved. Fitting of
the PL energies reveals the electron and hole $g$-factors. Other
(broad) spectral features correspond to PL of excitons charged
with more than one electron.}}
\end{figure}

The sample gate bias $V_{\rm{Gate}}$ is controlled by the output
of an arbitrary waveform generator connected through a 1.9~MHz low
pass filter. During the dark evolution time $T_{\rm{Dark}}$ the
bias can be set to an arbitrary value. For an empty dot regime
(0$e$) we use large reverse bias $V_{\rm{Gate}}=-1.3$~V. The bias
corresponding to 1$e$ Coulomb blockade is found by measuring the
bias dependence of $\varGamma_{\rm{N}}(V_{\rm{Gate}})$ such as
shown in Supplementary Fig.~\ref{Fig:SG1NvsBias}. In agreement
with the previous studies on InGaAs QDs
\citep{SupLatta2011,SupGillard2021} we observe tunnelling peaks
(at $\approx0.43$~V and $\approx0.6$~V), where the electron Fermi
reservoir energy matches the QD charging energy. A bias at the
middle of the Coulomb valley between the peaks, 0.517~V in this
case, is used to charge the QD with one electron (1$e$).
Supplementary Fig.~\ref{Fig:SG1NvsBias} shows that when the QD is
charged with two electrons (2$e$) the NSR rate is identical to the
0$e$ case, confirming that the NSR acceleration produced by the
single electron (1$e$) is related to its spin.

\begin{figure}
\includegraphics[width=0.5\linewidth]{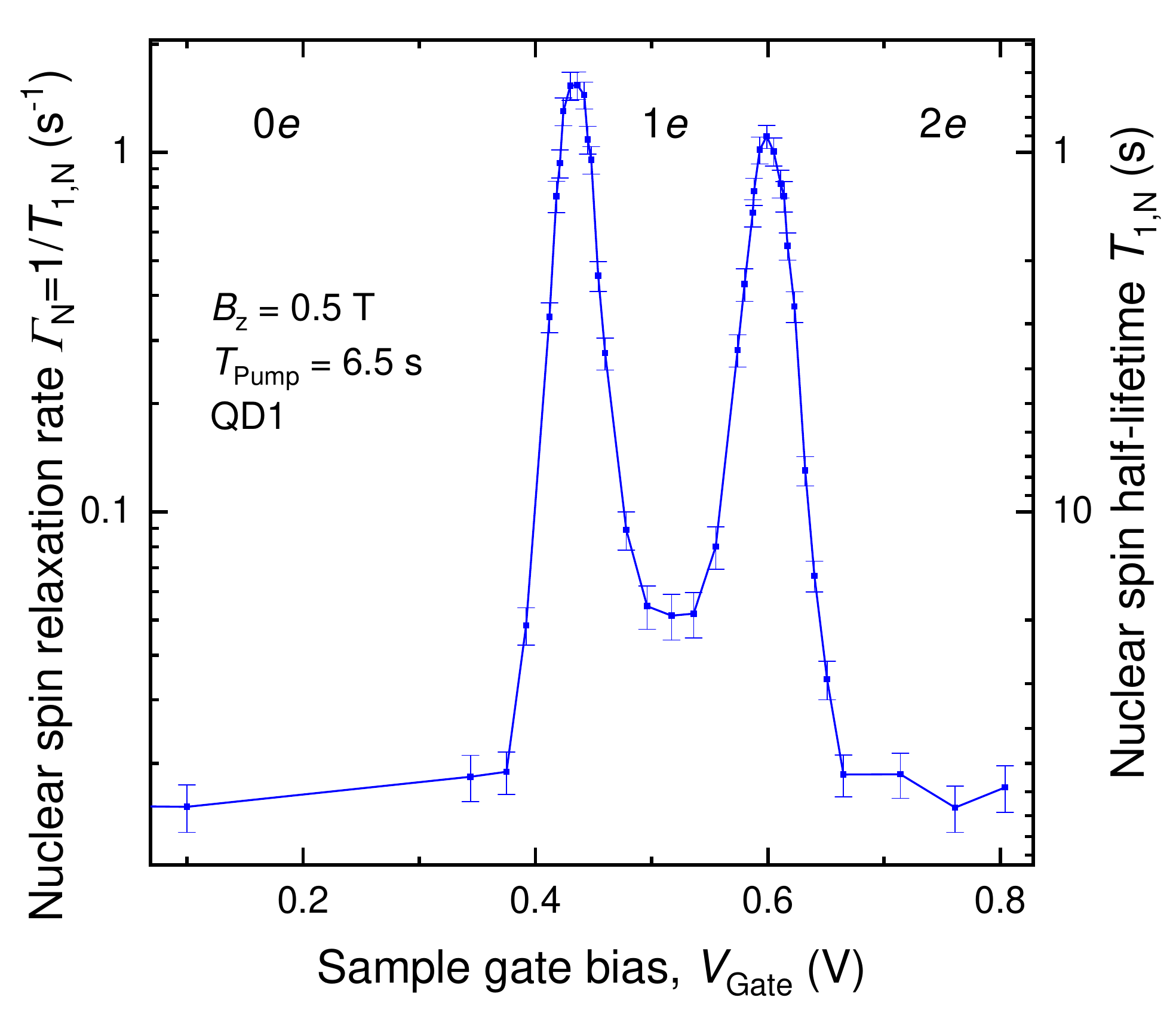}
\caption{\label{Fig:SG1NvsBias} Bias dependence of the nuclear
spin relaxation rate $\varGamma_{\rm{N}}$.}
\end{figure}

Optical pump and probe pulses are formed by mechanical shutters
with a switching time of a few milliseconds. Under certain regimes
in $B_{\rm{z}}$ and $V_{\rm{Gate}}$ this is comparable to the
nuclear spin relaxation times $T_{\rm{1,N}}$. However, the
relaxation time in an empty QD (0$e$) is always considerably
longer. Thus we keep QD under 0$e$ bias during the shutter
switching times and the dark time $T_{\rm{Dark}}$ is implemented
by pulsing the gate bias to the target dark-state value
$V_{\rm{Gate}}$ for a duration $T_{\rm{Dark}}$.

\RedText{Nuclear magnetic resonance (NMR) characterization
(Fig.~2(e) of the main text) is conducted using inverse NMR method
\citep{SupChekhovich2012} which enhances the signal for $I>1/2$
spins and improves the signal to noise ratio. In this method the
nuclei are depolarized by a weak RF field, whose spectral profile
is a broadband frequency comb with a narrow gap of width
$w_{\rm{gap}}$. The value of $w_{\rm{gap}}$ controls the balance
between NMR amplitude and spectral resolution. In an empty QD
(0$e$), NMR spectra of As and Ga measured with
$w_{\rm{gap}}=6$~kHz consist of well-resolved quadrupolar-split
triplets, consistent with previous observations for similar QD
structures \citep{SupUlhaq2016,SupChekhovich2018}. The spin of a
single electron (1$e$) leads to inhomogeneous Knight shifts
comparable to the quadrupolar splitting. As a result, the NMR
triplets are no longer resolved (solid circles in Fig.~2(e) of the
main text). Moreover, the electron spin lifetime, which is on the
order of milliseconds in the studied QDs, is much shorter than the
radio frequency burst (typically 0.18~s), and the average electron
spin polarization is therefore close to 0. Each nucleus then
experiences both positive and negative Knight shifts during the RF
burst. These dynamic spectral shifts disrupt the enhancement of
the inverse NMR method: for example, if a nuclear spin transition
fits into the RF spectral gap $w_{\rm{gap}}$ under one sign of the
Knight shift, it may be moved out of the gap and into resonance
with the depolarizing RF field under the opposite Knight shift. As
a result, the NMR spectrum amplitude is reduced in the 1$e$
measurement. By varying the gap width $w_{\rm{gap}}$, we find that
a spectrum with a reasonable signal to noise ratio is obtained at
$w_{\rm{gap}}=70$~kHz, as shown by the circles in Fig.~2(e) of the
main text. Although the deterioration of the inverse NMR method
precludes an accurate measurement of the NMR lineshape in presence
of the electron, the overall width $\approx50$~kHz of the
resonance still provides a valid order-of-magnitude estimate of
the Knight shifts experienced by the nuclear spins in the QD. More
sophisticated measurements, using pulsed NMR (to be reported
separately elsewhere) confirm this rough estimate based on inverse
NMR measurement.}

\begin{figure}
\includegraphics[width=0.8\linewidth]{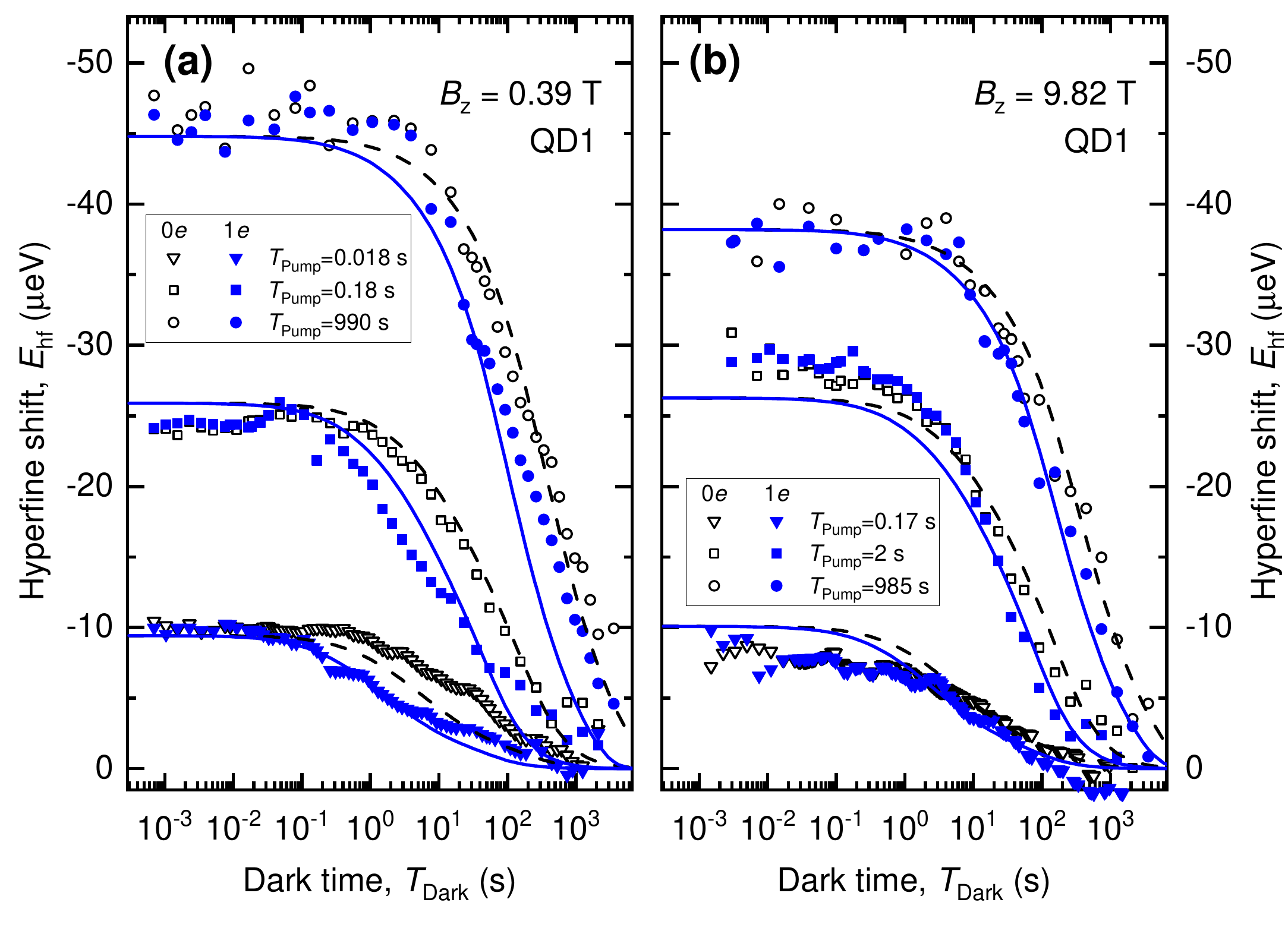}
\caption{\label{Fig:SNSR} (a) Dark time dependence of the
hyperfine shift $E_{\rm{hf}}$, which probes average nuclear spin
polarization weighted by the QD electron density
$|\psi_{\rm{e}}|^2$. Nuclear spin decay is measured (symbols) at
$B_{\rm{z}}=0.39$~T for different pumping times $T_{\rm{Pump}}$
while keeping QD empty (0$e$, open symbols) or electron-charged
(1$e$, solid symbols) during the dark time. Lines show numerical
solution of the spin diffusion equation. (b) Same as (a) but for
$B_{\rm{z}}=9.82$~T.}
\end{figure}

The QD NSR curves measured at $B_{\rm{z}}=0.39$~T and shown in
Fig.~3(b) of the main text are reproduced in Supplementary
Fig.~\ref{Fig:SNSR}(a), together with the similar measurements
carried out at high magnetic field $B_{\rm{z}}=9.82$~T and shown
in Supplementary Fig.~\ref{Fig:SNSR}(b). Similar to the low
fields, at $B_{\rm{z}}=9.82$~T shorter optical pumping time
$T_{\rm{Pump}}$ results in faster NSR through spin diffusion.
However, the acceleration of NSR in presence of a single electron
(1$e$) is less pronounced at high magnetic field, owing to the
reduction of the hyperfine-mediated nuclear-nuclear spin
interaction. It is also worth noting that the optical spin pumping
becomes slower at high magnetic field. While
$T_{\rm{Pump}}=0.018$~s at $B_{\rm{z}}=0.39$~T is sufficient to
achieve $\approx1/4$ of the steady state nuclear spin
polarization, it takes an order of magnitude longer
$T_{\rm{Pump}}=0.17$~s to reach the same $\approx1/4$ level at
$B_{\rm{z}}=9.82$~T. This difference limits the shortest
$T_{\rm{Pump}}$ for which NSR dynamics can be measured at high
magnetic field, as can be seen in Fig.~3(d) of the main text.

\begin{figure}
\includegraphics[width=0.5\linewidth]{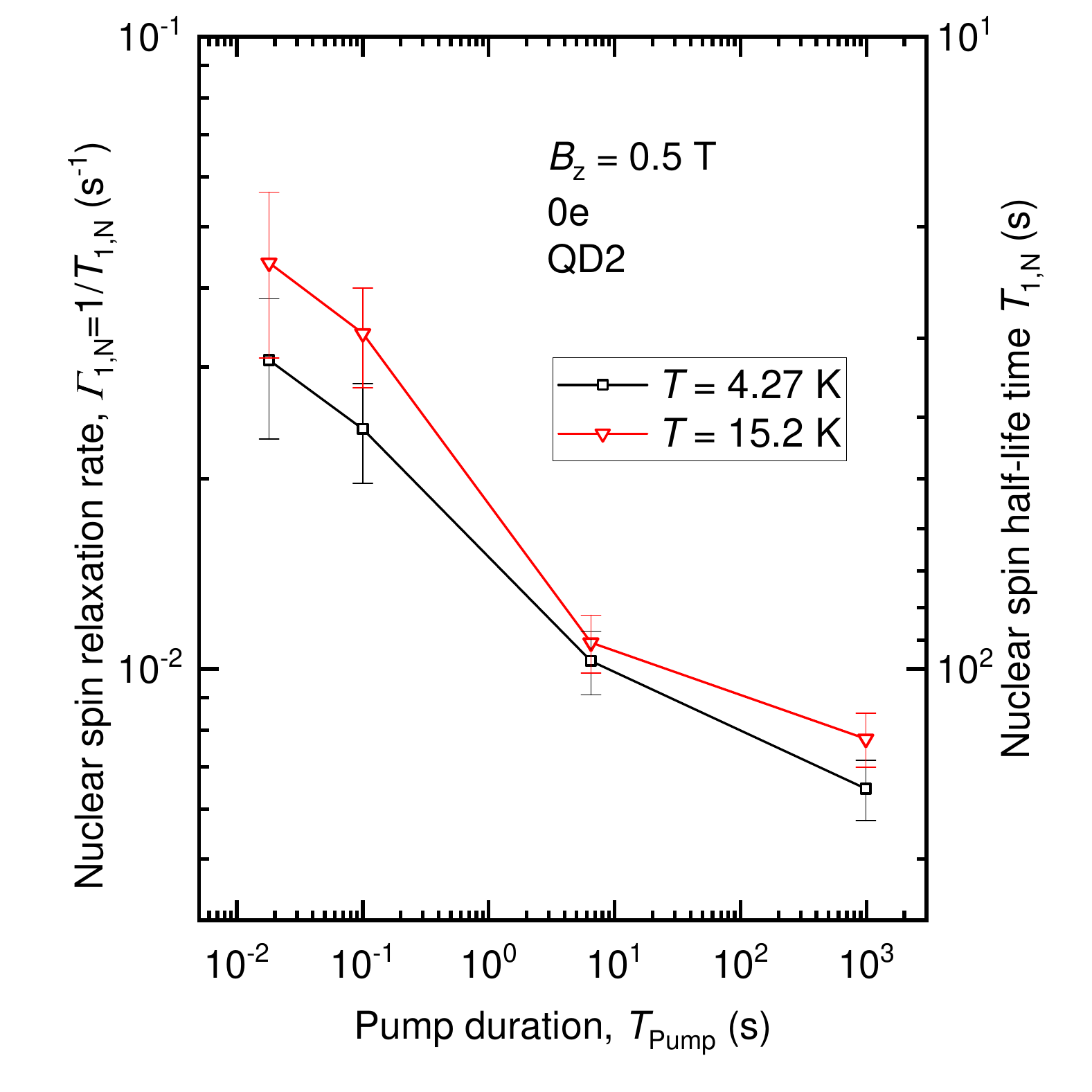}
\caption{\label{Fig:DiffTDep} Fitted QD nuclear spin half-life
times $T_{\rm{1,N}}$ (right scale) and corresponding NSR rates
$\varGamma_{\rm{N}}=1/T_{\rm{1,N}}$ (left scale) measured for
different pumping times $T_{\rm{Pump}}$ at $B_{\rm{z}}=0.5$~T.
Experiments are conducted at base sample temperature ($T=4.27$~K,
squares) and an elevated temperature ($T=15.2$~K, triangles).}
\end{figure}

The experiments presented in the main text are conducted at the
cryostat base temperature, measured with a resistive sensor to be
$T\approx4.27$~K. Additional measurements, similar to those shown
in Fig.~3(c,d) of the main text, have been conducted on an empty
QD (0$e$) at an elevated temperature $T=15.2$~K and are shown in
Supplementary Fig.~\ref{Fig:DiffTDep}. We find that at high
temperature the relaxation rate follows the same trend of
reduction at short pumping times $T_{\rm{Pump}}$, consistent with
NSR dominated by spin diffusion. In case of a pure spin diffusion
driven by nuclear dipole-dipole interactions, one would expect the
rate to be independent of the temperature. From Supplementary
Fig.~\ref{Fig:DiffTDep} we find that for any given $T_{\rm{Pump}}$
the relaxation is slightly accelerated at $T=15.2$~K. One
possibility is the temperature dependence of the optical nuclear
spin pumping process \citep{SupUrbaszek2007} creating different
spatial distributions of the nuclear spin polarization for the
same $T_{\rm{Pump}}$. Contribution of the temperature-dependent
non-diffusion mechanisms, such as two-phonon quadrupolar
relaxation \citep{SupMieher1962,SupMcNeil1976}, is also possible.

\section{First principle estimate of the G{\lowercase{a}}A{\lowercase{s}} nuclear spin diffusion coefficient}
\label{sec:DEstim}

In the absence of free electrons, nuclear spin diffusion is driven
by the dipole-dipole magnetic nuclear spin interaction. The total
dipole-dipole Hamiltonian term is a sum of pairwise couplings:
\begin{align}
\mathcal{H}_{\rm{DD}}=\sum_{i<j}b_{i,j}\left(2\hat{I}_{{\rm{z}},i}\hat{I}_{{\rm{z}},j}- \hat{I}_{{\rm{x}},i}\hat{I}_{{\rm{x}},j} - \hat{I}_{{\rm{y}},i}\hat{I}_{{\rm{y}},j} \right){\rm{,}}\nonumber\\
b_{i,j}=\frac{\mu_0}{4\pi}\frac{\gamma_{i}\gamma_{j}}{2}\frac{1-3\cos^2{\theta_{i,j}}}{r_{i,j}^3},\label{Eq:HDD}
\end{align}
where $\mu_0=4\pi\times 10^{-7}\;{\rm{N A}}^{-2}$ is the magnetic
constant and $r_{i,j}$ denotes the length of the vector, which
forms an angle $\theta$ with the static magnetic field direction
($z$) and connects the two spins $i$ and $j$. The typical
magnitude of the interaction constants for the nearby nuclei in
GaAs is $\max{(|b_{j,k}|)}/h\approx100$~Hz. The Hamiltonian of
Eq.~\ref{Eq:HDD} has been truncated to eliminate all spin
non-conserving terms, which is justified for static magnetic
fields exceeding $\gtrsim1$~mT, as used in this work. The
evolution of a large nuclear spin ensemble can be described in
terms of spin diffusion with coefficient $D$. In crystalline
solids the nuclear spin diffusion coefficient $D$, is a rank-2
tensor which can be calculated from the first principles using
density matrix approach \citep{SupLowe1967} or the method of
moments \citep{SupRedfield1968,SupRedfield1969}. The calculation
involves a somewhat lengthy evaluation of the various lattice
sums. Here we use a more recent version of the method of moments
from Ref.~\citep{SupButkevich1988}. We re-evaluate numerically the
sums of Eqns.~8 and 10 from Ref.~\citep{SupButkevich1988} using an
FCC lattice of 6859 spins. Our results are in good agreement with
the those derived for 1330 neighboring spins
previously~\citep{SupButkevich1988}. We find the following values
for the diagonal components of $D$:
$D_{xx}=D_{yy}\approx0.2594\frac{\mu_0}{4\pi}\frac{\hbar\gamma^2}{a_0}\rho^{1/3}$
and
$D_{zz}\approx0.3289\frac{\mu_0}{4\pi}\frac{\hbar\gamma^2}{a_0}\rho^{1/3}$,
where $\hbar$ is the reduced Planck's constant,
$a_0\approx0.565$~nm is the GaAs lattice constant, and $\gamma$ is
the nuclear gyromagnetic ratio. Here we use the coordinate system
aligned with the cubic crystal axes $x\parallel[100]$,
$y\parallel[010]$, $z\parallel[001]$, and the strong magnetic
field is parallel to the $z$ direction. We have also introduced
the correction factor $\rho^{1/3}$ to account for the increase of
the average internuclear distance for the isotope whose abundance
$\rho$ is less then unity.

In case of arsenic, $^{75}$As is the only stable isotope, so that
$\rho=1$. For gallium isotopes we have the natural abundances
$\rho=0.601$ and $\rho=0.399$ for $^{69}$Ga and $^{71}$Ga,
respectively. The gyromagnetic ratios $\gamma$ are known
\citep{SupHARRIS2002458} and, since we approximate the spin
diffusion as a one-dimensional process along the sample growth
direction $z$, we are interested in the $D_{zz}$ component of the
tensor. Substituting the numerical values we find
$D_{zz}\approx13$, 21, 30~nm$^2$~s$^{-1}$ for $^{75}$As, $^{69}$Ga
and $^{71}$Ga, respectively. The experiments presented in this
work do not resolve between spin diffusion of the individual
isotopes. As a simple approximation we can treat the observed NSR
dynamics as a result of spin diffusion within one type of nuclei
but with a weighted average diffusion constant. We use as weights
the relative contributions of the isotopes to the optically
measured hyperfine shift $E_{\rm{hf}}$. From the previous studies
of the similar QDs \citep{SupChekhovich2017} these contributions
are estimated as 0.49, 0.28 and 0.23 for $^{75}$As, $^{69}$Ga and
$^{71}$Ga, respectively, from where the average diffusion
coefficient is approximated as $D_{zz}\approx19$~nm$^2$.

\section{Numerical simulation of nuclear spin diffusion coefficient and additional analysis}
\label{sec:DiffNum}

The spatiotemporal evolution of the nuclear spin polarization
degree $P _{\rm{N}}(t,z)$ is modeled by solving the partial
differential spin diffusion equation
\begin{align}
\frac{\partial P_{\rm{N}}(t,z)}{\partial
t}=D(t)\frac{{\partial}^{2} P_{\rm{N}}(t,z)}{\partial z^2} +
w(t)|\psi_{\rm{e}}(z)|^2 (P_{\rm{N},0}-P
_{\rm{N}}(t,z)),\label{Eq:SDiffEq}
\end{align}
where the last term describes optical nuclear spin pumping with a
rate proportional to electron density $|\psi_{\rm{e}}(z)|^2$ and
the time-dependent factor $w(t)$ equal to 0 or $w_0$ when optical
pumping is off or on, respectively. The spin diffusion coefficient
$D(t)$ also takes two discrete values $D_{\rm{Dark}}$ or
$D_{\rm{Pump}}$ when optical pumping is off or on, respectively.
$P_{\rm{N},0}$ is a steady state nuclear spin polarization degree
that optical pumping would generate in the absence of spin
diffusion. At each time point we assume the same diffusion
coefficient $D$ across the entire structure. The equation
describes a one-dimensional problem where diffusion can take place
only along the $z$ coordinate so that the nuclear spin
polarization degree $P_{\rm{N}}$ does not depend on $x$ or $y$.
The GaAs QD layer is modeled by taking a Gaussian profile for the
electron density
$|\psi_{\rm{e}}(z)|^2\propto2^{-\left(\frac{z-z_0}{h_{\rm{QD}}/2}\right)^2}$,
where $h_{\rm{QD}}$ is the full width at half maximum of the
$|\psi_{\rm{e}}(z)|^2$ function and the center of the QD is set to
be $z_0=0$. We use Dirichlet boundary condition $P_{\rm{N}}=0$ to
model fast nuclear spin depolarization in presence of the free
carriers both in the $n$- and $p$-type doped layers. The boundary
coordinates where Dirichlet conditions are enforced are chosen to
match the actual sample structure as described
in~\ref{sec:Sample}. \RedText{We note that the hyperfine
interaction of the valence band holes is approximately 10 times
weaker than for the conduction band electrons
\citep{SupChekhovich2013NatPhys}. Moreover, the $p$-type layer is
approximately 10 times further away from the QDs than the $n$-type
layer. As a result the dynamics of the nuclear spin polarization
at the QD are dominated by the $n$-type layer, while the exact
boundary condition at the $p$-type layer is less crucial,
justifying the use of the Dirichlet condition at both doped
layers.}

\begin{figure}
\includegraphics[width=0.8\linewidth]{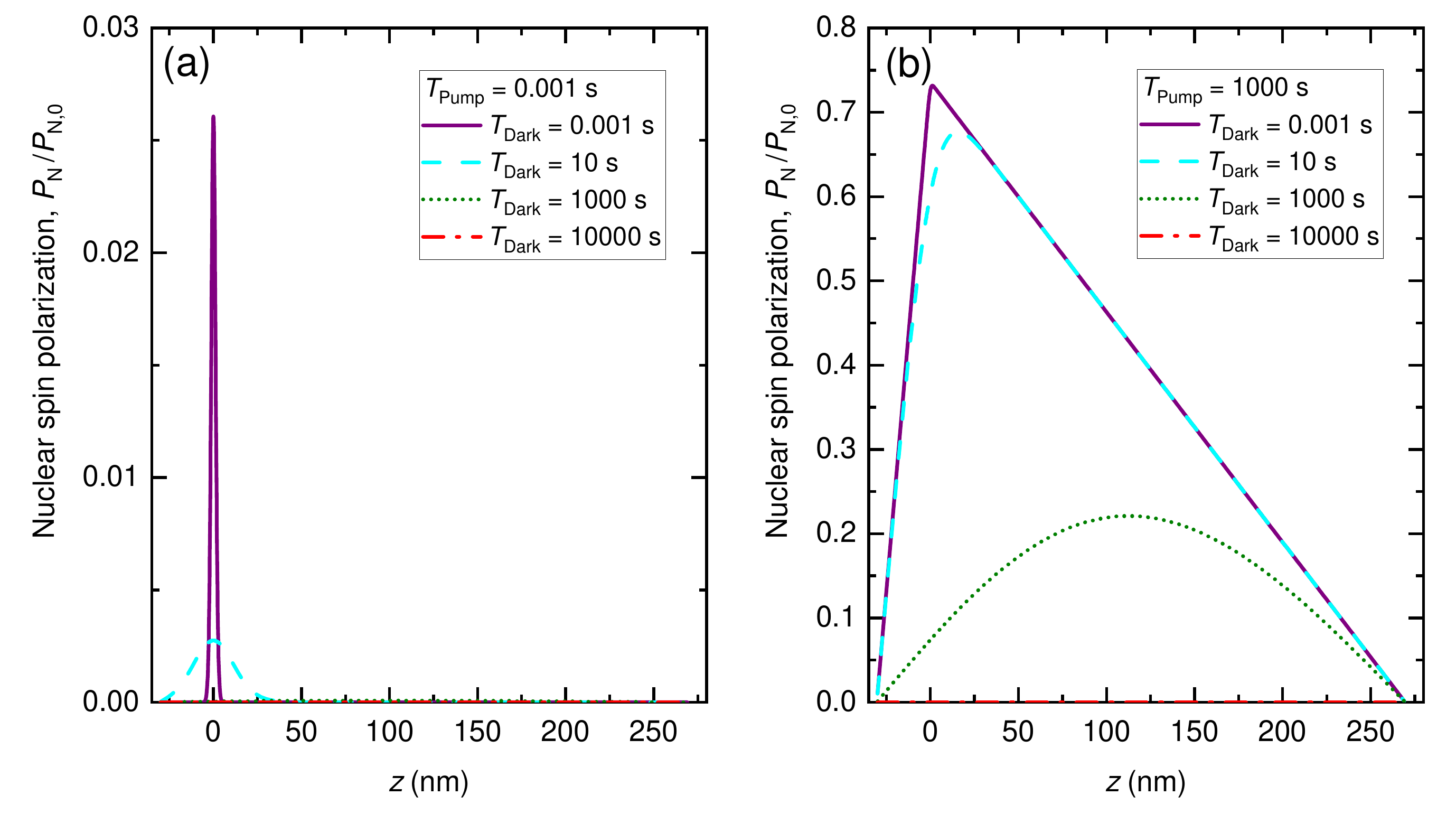}
\caption{\label{Fig:SDiffzt} Calculated normalized nuclear spin
polarization as a function of the $z$ coordinate at different
$T_{\rm{Dark}}$. (a) Calculations for $T_{\rm{Pump}}=1$~ms. (b)
Same calculations for $T_{\rm{Pump}}=1000$~s.}
\end{figure}

Supplementary Eq.~\ref{Eq:SDiffEq} is solved numerically using the
method of lines implemented in \textsc{wolfram mathematica} 12.0.
The initial condition is taken to be $P_{\rm{N}}=0~\forall~z$,
which models the result of the RF Erase pulse at the start of each
measurement cycle. Optical nuclear spin pumping starts at
$t=-T_{\rm{Pump}}$ and the equation is solved until $t=0$ with
$D=D_{\rm{Pump}}$ and $w=w_0$. At $t=0$ optical pumping is
switched off by setting $w=0$ and the equation is solved until
$t=T_{\rm{Dark}}$ with $D=D_{\rm{Dark}}$. Supplementary
Fig.~\ref{Fig:SDiffzt} shows the calculated spatial profiles of
the final nuclear spin polarization $P_{\rm{N}}(T_{\rm{Dark}},z)$
normalized by its steady-state value $P_{\rm{N},0}$. The results
are shown for several $T_{\rm{Dark}}$ values in case of a short
pumping [(a), $T_{\rm{Pump}}=1$~ms] and long pumping [(b),
$T_{\rm{Pump}}=1000$~s]. Short pumping results in a
small-magnitude ($P_{\rm{N}}\ll P_{\rm{N},0}$) spatially-narrow
nuclear spin polarization, which quickly dissipates at $t>0$. By
contrast, long pumping leads to a steady-state spatial
distribution where polarization peaks at the quantum dot
coordinate $z=0$ and reduces linearly towards the doped layers
which act as nuclear spin polarization sinks. Interestingly, this
calculation predicts that the maximum polarization $P_{\rm{N},0}$
is not achieved because of the diffusion towards the doped layers,
especially the closely located $n$-type layer at $z<0$.

In order to compare simulations with the experimental results the
final spatial distribution $P_{\rm{N}}(T_{\rm{Dark}},z)$ is
multiplied by $|\psi_{\rm{e}}(z)|^2$ and integrated over $z$. This
way we reproduce the optical probing of the nuclear spin
polarization, where the measured hyperfine shift $E_{\rm{hf}}$ is
effectively weighted by the electron envelope wavefunction density
$|\psi_{\rm{e}}(z)|^2$. The simulated hyperfine shift is then
derived as $E_{\rm{hf}}=E_{\rm{Z}^{\rm{X^-}}}+A I P_{\rm{N}}$,
where $I$ is the nuclear spin number, $A$ is the hyperfine
constant and $E_{\rm{Z}^{\rm{X^-}}}$ is the trion Zeeman splitting
in the absence of nuclear spin polarization. We then use a
differential evolution algorithm to vary the parameters such as
$D^{(ne)}_{\rm{Dark}}(B_{\rm{z}})$, $w_0(B_{\rm{z}})$ and
$D_{\rm{Pump}}(B_{\rm{z}})$ and fit the simulated $E_{\rm{hf}}$
dynamics to the entire experimental datasets of
$E_{\rm{hf}}(T_{\rm{Pump}},T_{\rm{Dark}})$ measured at
$B_{\rm{z}}=0.39$ and $9.82$~T  for empty ($n=0$) and charged
($n=1$) QD states. As discussed in the main text the best-fit
diffusion coefficients in the dark are
$D^{(0e)}_{\rm{Dark}}=2.2^{+0.7}_{-0.5}$~nm$^2$~s$^{-1}$
independent of magnetic field,
$D^{(1e)}_{\rm{Dark}}(9.82~{\rm{T}})=4.7^{+1.2}_{-1.0}$~nm$^2$~s$^{-1}$
and
$D^{(1e)}_{\rm{Dark}}(0.39~{\rm{T}})=7.7\pm1.9$~nm$^2$~s$^{-1}$.
For spin diffusion coefficients under optical pumping we find
significantly larger values
$D^{(1e)}_{\rm{Pump}}(9.82~{\rm{T}})=96^{+44}_{-28}$~nm$^2$~s$^{-1}$
and
$D^{(1e)}_{\rm{Pump}}(0.39~{\rm{T}})=850^{+240}_{-220}$~nm$^2$~s$^{-1}$.
This can be ascribed to the spectrally broad fluctuations of the
optically generated electron spins which provide coupling between
the distant nuclear spins, thus facilitating the diffusion. The
other best-fit parameters are
$w_0(0.39~{\rm{T}})=37^{+7}_{-5}$~s$^{-1}$,
$w_0(9.82~{\rm{T}})=5.7^{+1.1}_{-0.9}$~s$^{-1}$ and
$h_{\rm{QD}}=2.1^{+0.3}_{-0.2}$~nm. Previous studies on
GaAs/AlGaAs QDs emitting at a similar wavelength estimated that
0.92 of the electron density resides in the GaAs layer
\citep{SupChekhovich2017}, whose full width can then be estimated
as
$h_{\rm{QD}}\frac{{\rm{erf}}^{-1}(0.92)}{\sqrt{\ln(2)}}=3.2^{+0.4}_{-0.3}$~nm.
This best-fit value somewhat underestimates the true QD thickness
in $z$ direction, but is within the range bounded by the QW
thickness (2.1~nm) and the maximum QD thickness ($\approx9$~nm)
estimated from the nanohole depth in AFM. The spatial profiles
shown in Supplementary Fig.~\ref{Fig:SDiffzt} are calculated with
the best-fit parameters for the 1$e$ case at $B_{\rm{z}}=0.39$~T.

\begin{figure}
\includegraphics[width=0.8\linewidth]{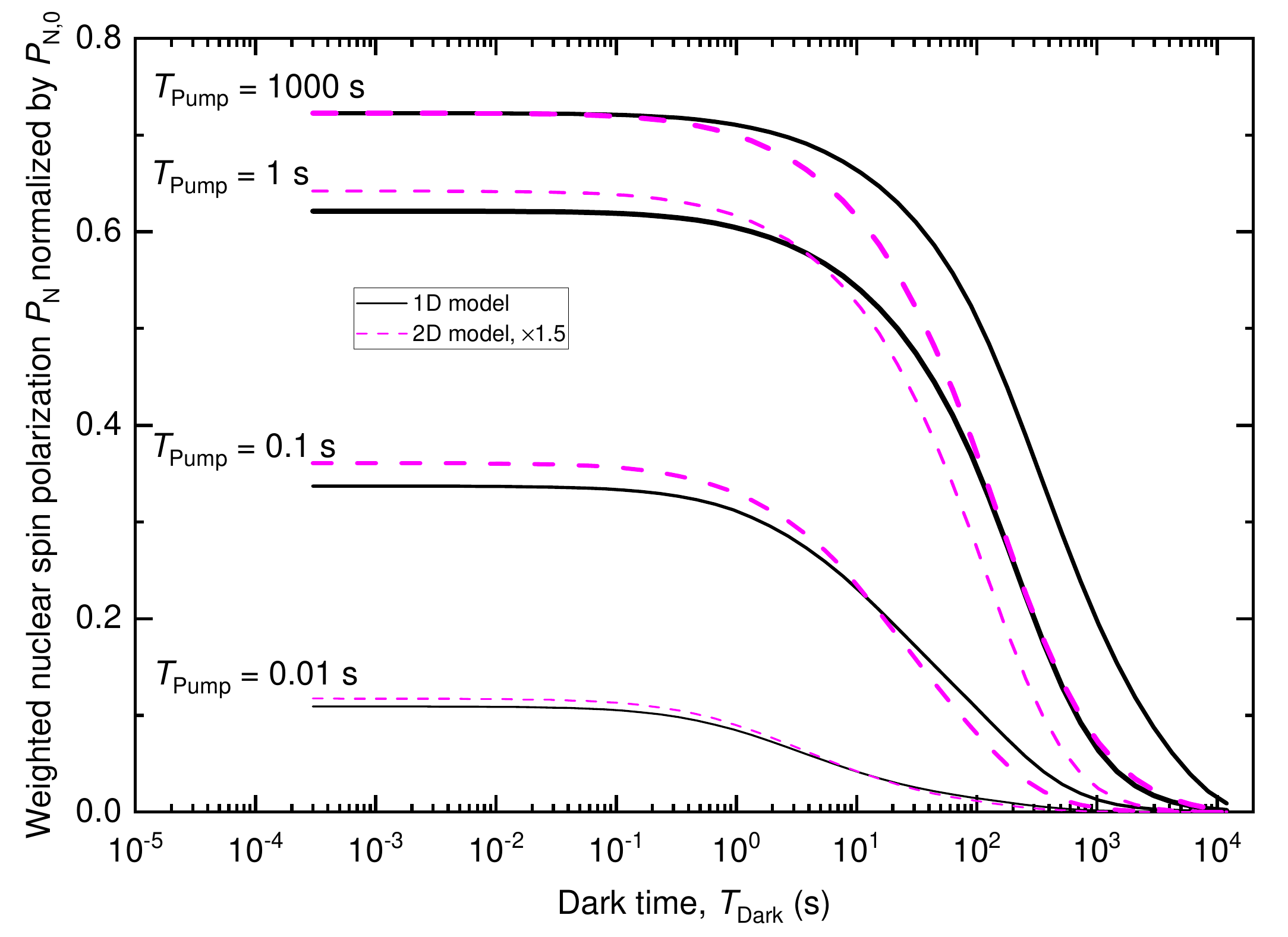}
\caption{\label{Fig:S1D2D} Numerically simulated nuclear spin
polarization degree $P_{\rm{N}}$ weighted by the electron envelope
wavefunction density $|\psi_{\rm{e}}(z)|^2$ and normalized by the
maximum nuclear spin polarization $P_{\rm{N},0}$ in the absence of
spin diffusion. The weighted polarization is plotted as a function
of the dark time $T_{\rm{Dark}}$ for different $T_{\rm{Pump}}$.
The results are shown for the case of one dimensional diffusion
(1D, solid lines) and two dimensional diffusion (2D, dashed lines,
$P_{\rm{N}}$ values multiplied by 1.5).}
\end{figure}

The use of the one dimensional spin diffusion model is motivated
by the large aspect ratio of the QD: the diffusion proceeds
predominantly along the direction of the strongest gradient in the
nuclear spin polarization degree, which is the growth $z$
direction. For numerical simulations, one dimensional model is
also advantageous as it requires significantly less computational
resources than a full three dimensional diffusion model. In order
to evaluate the limitations of the one dimensional model we run a
simulation of a two dimensional diffusion problem, where the
equation now reads:
\begin{align}
\frac{\partial P_{\rm{N}}(t,x,z)}{\partial
t}=D(t)\left(\frac{{\partial}^{2} P_{\rm{N}}(t,x,z)}{\partial
x^2}+\frac{{\partial}^{2} P_{\rm{N}}(t,x,z)}{\partial z^2}\right)
+ w(t)|\psi_{\rm{e}}(x,z)|^2 (P_{\rm{N},0}-P
_{\rm{N}}(t,x,z))\label{Eq:SDiffEq2D}
\end{align}
The electron density is taken to be
$|\psi_{\rm{e}}(x,z)|^2\propto2^{-\left(\frac{x-x_0}{d_{\rm{QD}}/2}+\frac{z-z_0}{h_{\rm{QD}}/2}\right)^2}$,
where $d_{\rm{QD}}$ is a full width at half maximum diameter of
the QD, which we set to $d_{\rm{QD}}=47$~nm in order to match the
0.92 electron wavefunction density in a QD with a full diameter of
$70~$nm. The same $|\psi_{\rm{e}}(x,z)|^2$ is used to calculate
the weighted nuclear spin polarization degree, emulating the
optical probing of the QD hyperfine shift $E_{\rm{hf}}$. The
computational domain is limited to $|x|<700$~nm and we implement
the additional Dirichlet boundary condition
$P_{\rm{N}}(x=\pm700~{\rm{nm}})=0$.

Supplementary Fig.~\ref{Fig:S1D2D} shows the simulated QD NSR
dynamics in a one dimensional (1D, solid lines) and two
dimensional (2D, dashed lines) cases, following nuclear spin
pumping with different durations $T_{\rm{Pump}}$. One difference
in the resulting dynamics is the lower weighted nuclear spin
polarization degree within the QD volume in the 2D case.
Consequently, all the 2D-case $P_{\rm{N}}$ values in Supplementary
Fig.~\ref{Fig:S1D2D} have been multiplied by 1.5, to simplify
comparison with the 1D case. At short $T_{\rm{Pump}}\leq0.1$~s QD
nuclear spin polarization decays on the same timescale both in 1D
and 2D cases. This is expected: the spatial profile of the nuclear
spin polarization produced by short pumping in 2D case is
proportional to $\propto|\psi_{\rm{e}}(x,z)|^2$, so that the
subsequent diffusion in the dark proceeds predominantly along the
direction of the highest gradient (the growth $z$ direction),
making diffusion essentially one dimensional. By contrast, long
pumping $T_{\rm{Pump}}\geq0.1$~s in a 2D model makes the
polarization profile more isotropic in the $xz$ plane (for an
unbounded problem at $T_{\rm{Pump}}\rightarrow\infty$ the
polarization will tend to a profile with circular contour lines in
the $xz$ plane). In other words, after long pumping the system
``forgets'' the initial profile $\propto|\psi_{\rm{e}}(x,z)|^2$ of
the QD pumping source. The subsequent diffusion in the absence of
pumping (i.e. in the dark) is controlled by the dimensionality of
the unpolarized space, and is seen to be faster in the 2D case.
From this additional results we conclude that one dimensional
model captures the key aspects of QD NSR dynamics, such as slower
relaxation following long optical nuclear spin pumping. However,
some quantitative discrepancies arise, especially at long
$T_{\rm{Pump}}$, where dimensionality affects the diffusion
dynamics. Such discrepancies are likely to introduce systematic
errors in the best fit values of the spin diffusion coefficient
$D$. However, in a real QD system $D$ is not constant, and the
approximate nature of the spin diffusion concept itself entails a
range of systematic errors. This justifies the use of a simplified
one dimensional model to describe our experimental results.


\end{document}